\documentclass{SciPost}

\usepackage{amsthm}
\usepackage{relsize}
\usepackage{physics}
\usepackage{bbold}

\usepackage[utf8]{inputenc} 
\usepackage[T1]{fontenc} 	
\usepackage[english]{babel} 

\usepackage[bitstream-charter]{mathdesign}
\usepackage{geometry} 		
\usepackage{amsmath} 		
\usepackage{mathtools} 		
\usepackage{float} 			
\usepackage{graphicx} 		
\usepackage{tabularx} 		
\usepackage{booktabs} 		
\usepackage{color, xcolor} 	
\usepackage{pdfpages} 		
\usepackage{extarrows} 		
\usepackage{multirow} 		
\usepackage{multicol} 		
\usepackage{enumitem} 		
\usepackage{xspace} 		
\usepackage{stackrel} 		
\usepackage{tikz} 			
\usepackage{braket} 		
\usepackage{bm} 			
\usepackage{tensor} 		
\usepackage{slashed} 		
\usepackage{lastpage} 		
\usepackage{cite} 			
\usepackage[normalem]{ulem} 
\usepackage{fontawesome} 	
\usepackage{tocloft} 		
\usepackage{titlesec} 		
\usepackage{doi} 			
\usepackage{hyperref} 		
\usepackage[most]{tcolorbox} 					
\usepackage[nameinlink, capitalize]{cleveref} 	
\usepackage[nottoc, notlot, notlof]{tocbibind} 	
\usepackage[ruled, vlined]{algorithm2e} 		
\usepackage{makecell}
\usepackage[makeroom]{cancel}
\usepackage{feynmf}
\usepackage{aas_macros}          
\usepackage{pifont} 
\newcommand{\cmark}{\ding{51}}%
\newcommand{\xmark}{\ding{55}}%

\fancypagestyle{SPstyle}{
\fancyhf{}
\lhead{\colorbox{scipostblue}{\bf \color{white} ~SciPost Physics }}
\rhead{{\bf \color{scipostdeepblue} ~Submission }}

\fancyfoot[C]{\textbf{\thepage}}
}

\makeatletter
\def\BState{\State\hskip-\ALG@thistlm}
\makeatother

\makeatletter
\@ifundefined{pdfoutput}{}{\DeclareGraphicsRule{*}{mps}{*}{}}
\makeatother

\makeatletter
\DeclareRobustCommand*{\bfseries}{%
   \not@math@alphabet\bfseries\mathbf
   \fontseries\bfdefault\selectfont
   \boldmath
}
\makeatother

\hypersetup{
	pdftitle={SKATR: A Self-Supervised Summary Transformer for SKA},
	pdfauthor={Ore et al.},
	colorlinks=true, 			
	linkcolor={red!50!black}, 	
	citecolor={blue!50!black}, 	
	urlcolor={blue!80!black} 	
} 

\DeclareSymbolFont{usualmathcal}{OMS}{cmsy}{m}{n}
\DeclareSymbolFontAlphabet{\mathcal}{usualmathcal}

\SetArgSty{textnormal}
\SetKwComment{Comment}{{\small\#}~}{}
\SetCommentSty{mycommfont}

\setitemize{itemsep=0pt, parsep=0pt} 				
\setenumerate{itemsep=0pt, parsep=0pt} 				
\setitemize{itemsep=2pt,topsep=2pt,parsep=0pt,partopsep=0pt,leftmargin=*}
\setenumerate{itemsep=0pt,topsep=2pt,parsep=0pt,partopsep=0pt,labelindent=3pt,leftmargin=*}
\usepackage{amsmath}
 
\usepackage{amsthm} 		
\theoremstyle{definition}

\definecolor{Rcolor}{HTML}{E99595}
\definecolor{Gcolor}{HTML}{C5E0B4}
\definecolor{Gcolor_light}{HTML}{F1F8ED}
\definecolor{Bcolor}{HTML}{9DC3E6}
\definecolor{Ycolor}{HTML}{FFE699}
\definecolor{Pcolor}{HTML}{C8B7E1}
\definecolor{Ycolor_light}{HTML}{FFF7DE}

\usetikzlibrary{arrows, arrows.meta, shapes, positioning}

\tikzstyle{arrow} = [thick,-{Latex[scale=1.0]}, line width=0.2mm, color=black]
\tikzstyle{loss} = [circle, thick, rounded corners=0.3ex, minimum width=1.5cm, minimum height=1cm, text centered, align=center, inner sep=0, fill=white, font=\large, draw]
\tikzstyle{net} = [double arrow, double arrow head extend=0cm, double arrow tip angle=130, shape border rotate=90, inner sep=0, align=center, minimum width=2.1cm, minimum height=2.3cm, fill=Bcolor, draw,font=\large]
\tikzstyle{net_black} = [net, minimum height=2.5cm, fill=black]
\tikzstyle{expr} = [rectangle, rounded corners=0.3ex, minimum width=2.4cm, minimum height=1cm, text centered, align=center, inner sep=0, fill=Ycolor, font=\large, draw]

\DeclareSymbolFont{usualmathcal}{OMS}{cmsy}{m}{n}
\DeclareSymbolFontAlphabet{\mathcal}{usualmathcal}

\definecolor{red_cb}{HTML}{e41a1c}
\definecolor{blue_cb}{HTML}{377eb8}
\definecolor{green_cb}{HTML}{4daf4a}
\definecolor{purple_cb}{HTML}{984ea3}
\definecolor{orange_cb}{HTML}{ff7f00}

\definecolor{EmeraldGreen}{HTML}{1ea78d}
\definecolor{EnglishRed}{HTML}{b02427}
\hypersetup{colorlinks=true,urlcolor=EmeraldGreen,citecolor=EmeraldGreen,linkcolor=EnglishRed}

\newcommand\one{\leavevmode\hbox{\small1\normalsize\kern-.33em1}}

\newcommand{\ReLU}{\operatorname{ReLU}}

\newcommand{\Attention}{\operatorname{Attention}}

\newcommand{\mWDM}{\ensuremath{m_\text{WDM}}\xspace}
\newcommand{\Om}{\ensuremath{\Omega_\text{m}}\xspace}
\newcommand{\LX}{\ensuremath{L_\text{X}}\xspace}
\newcommand{\Tvir}{\ensuremath{T_\text{vir}}\xspace}

\newcommand{\loss}{\mathcal{L}} 	

\newcommand{\arXiv}[2][]{%
	\ifthenelse{\equal{#1}{}}%
	{\href{http://arxiv.org/abs/#2}{arXiv:#2}}%
	{\href{http://arxiv.org/abs/#2}{arXiv:#2~[#1]}}}

\def\slashchar#1{\setbox0=\hbox{$#1$}           
   \dimen0=\wd0                                 
   \setbox1=\hbox{/} \dimen1=\wd1               
   \ifdim\dimen0>\dimen1                        
      \rlap{\hbox to \dimen0{\hfil/\hfil}}      
      #1                                        
   \else                                        
      \rlap{\hbox to \dimen1{\hfil$#1$\hfil}}   
      /                                         
   \fi}

\newcommand{\tikznode}[2]{%
\ifmmode%
\tikz[remember picture,baseline=(#1.base),inner sep=0pt] \node (#1) {$#2$};%
\else
\tikz[remember picture,baseline=(#1.base),inner sep=0pt] \node (#1) {#2};%
\fi}

\def\mathswitchr#1{\relax\ifmmode{\mathrm{#1}}\else$\mathrm{#1}$\xspace\fi}
\def\mathswitch#1{\relax\ifmmode#1\else$#1$\xspace\fi}

\newcommand{\skatr}{\textsc{Skatr}\xspace}

\graphicspath{{./figures/}}
\begin{document}

\pagestyle{SPstyle}

\vspace*{-2.5em}
\hfill{}
\vspace*{0.5em}

\begin{center}{\Large \textbf{
      \textsc{Skatr}: A Self-Supervised Summary Transformer for SKA
}}\end{center}

\begin{center}
  Ayodele Ore\textsuperscript{1}, 
  Caroline Heneka\textsuperscript{1}, and
  Tilman Plehn\textsuperscript{1,2}
\end{center}

\begin{center}
{\bf 1} Institut für Theoretische Physik, Universität Heidelberg, Germany\\
{\bf 2} Interdisciplinary Center for Scientific Computing (IWR), Universit\"at Heidelberg,
Germany
\end{center}

\begin{center}
\today
\end{center}

\section*{Abstract}
{\bf
The Square Kilometer Array will initiate a new era of radio astronomy by allowing 3D imaging of the Universe during Cosmic Dawn and Reionization. Modern machine learning is crucial to analyze the highly structured and complex signal. However, accurate training data is  expensive to simulate, and supervised learning may not generalize. We introduce a self-supervised vision transformer, \skatr, whose learned encoding can be cheaply adapted for downstream tasks on 21cm maps. Focusing on regression and generative inference of astrophysical and cosmological parameters, we demonstrate that \skatr representations are maximally informative and that \skatr generalizes out-of-domain to differently-simulated, noised, and higher-resolution datasets.
}

\vspace{5pt}
\noindent\rule{\textwidth}{1pt}
\renewcommand{\baselinestretch}{0.92}\normalsize
\tableofcontents\thispagestyle{fancy}
\renewcommand{\baselinestretch}{1}\normalsize
\noindent\rule{\textwidth}{1pt}

\clearpage
\section{Introduction}
\label{sec:intro}

The Square Kilometer Array (SKA) is a discovery machine that has
recently seen first light.  The SKA-LOW part of the radio
interferometer is sensitive to the 21cm background fluctuations of
neutral hydrogen during the Epoch of Reionization (EoR) and Cosmic
Dawn (CD).
Its large-scale signal is sensitive to the thermal and cosmological
evolution of the Universe, making it a tracer also of primordial
sources of radiation and fundamental
physics~\cite{Heneka:2018kgn,LiuWP2019,Liu:2020JCAP,Berti:2022,Pourtsidou:2016,Modak:2022gol,Evoli:2014,List2020,Jones:FDM2021}.

The SKA will provide deep 3D-imaging of over 50$\%$ of the observable
Universe. This tomography of the large-scale structure will produce
data rates of several TB/s and archival data of up to hundreds of
PB/a~\cite{doi:10.1098/rsta.2019.0060}. At the same time, the SKA data
suffers from a complexity problem: foregrounds and systematics from
interferometric reconstruction cannot be modeled accurately for
synthetic data; simulations via hydrodynamical and radiative transfer
or approximate hydrodynamical (semi-numerical) simulations suffer from
model
uncertainties~\cite{Mes07,Mesinger2010,Hassan:2016,Heneka2018,Hutter:2021,Meriot:2023};
and an analytical optimal summary to construct a likelihood does not
exist.

Classical analyses use physics-motivated highly compressed
summaries. Analyses based on the power spectrum assume Gaussianity,
motivated by CMB assuming standard cosmology. This assumption breaks
down for the highly non-Gaussian SKA signal. Beyond-Gaussian
statistics such as bispectra and morphological diagnostics improve 
constraints and reduce bias for parameter inference from 21cm
intensity maps~\cite{Shimabukuro:2017,Majumdar:2018}; a picture which
might revert when faced with foregrounds, as these methods do not
generalize~\cite{Watkinson:2021}.

Modern machine learning opens a path toward optimality in
data-intensive analyses in fundamental physics and
cosmology~\cite{Boucaud:2020,Dvorkin:2022,Plehn:2022ftl,Villaescusa-Navarro:2022,Moriwaki:2023,Lanzieri:2024mvn,Makinen:2024xph},
including optimal compression for robust performance at downstream
tasks such as inference~\cite{Schosser:2024aic}.  Especially for SKA
data we need to bridge different simulators and assumptions on noise
and systematics, while remaining maximally
informative~\cite{Hartley:2023}. Supervised approaches may not
generalize well enough, and the size of realistic simulated datasets
is limited.

We propose to use self-supervised learning to train a
maximally-informative summary network that can easily be adapted for
downstream tasks without fine-tuning. Self-supervised approaches
based on contrastive learning or masking have been shown to generate
expressive representations, both in
vision~\cite{pmlr-v119-chen20j,NEURIPS2020_f3ada80d,Chen_2021_CVPR,Caron_2021_ICCV,bao2022beit,zhou2022image,pmlr-v162-baevski22a,Assran_2023_CVPR,bardes2024revisiting}
and fundamental physics~\cite{Dillon:2021gag,Golling:2024abg,Birk:2024knn,Leigh:2024ked}. Similarly,
self-supervised learning has been shown to aid data compression and
inference for galaxy surveys~\cite{Akhmetzhanova:2023hiy, 10.1093/mnras/stae1450}, including the use of pre-trained foundation models~\cite{2024arXiv240911175L}. Given the
large volume probed, systematics from radio interferometric
measurements, and the less known adequate summary statistics for 21cm
physics, we expect the benefits to be even larger for SKA data.

In this analysis, we focus on regression and inference of a set of astrophysical and cosmological parameters from SKA lightcones. We first establish that Vision Transformers~\cite{dosovitskiy2021an} are a suitable network architecture by comparing with the current CNN benchmark~\cite{Neutsch:2022hmv,Heneka:2023}. We then show that our self-supervised SKA Transformer~(\skatr) learns a near-lossless compression. In particular, a shallow MLP trained on frozen \skatr{} summaries matches the performance of a ViT trained from scratch, at much higher efficiency. Using datasets simulated at different resolutions, we find that \skatr generalizes well when faced with parameter information absent during the pre-training, as well as with instrumental and thermal noise and foreground avoidance. Moreover, \skatr generalizes better than a summary pre-trained with full supervision.

The remainder of the paper is organized as
follows. Section~\ref{sec:data} details our simulated datasets as well
as the transformations we employ for preprocessing and data
augmentation. In Section~\ref{sec:skatr} we introduce the network
architecture and self-supervised training strategy that comprise
\skatr. A series of results showing the benefit of \skatr are presented in
Section~\ref{sec:res}. In particular, we show that 
the fixed \skatr summary a) matches supervised benchmark networks in regression and inference tasks (Section~\ref{sec:res_skatr} and Section~\ref{sec:res_inf}), b) generalizes out-of-domain regarding noise and new parameter correlations present in higher resolution data (Section~\ref{sec:res_gen}), c) outperforms supervised baselines when data is limited (Section~\ref{sec:res_eff}), d) performs well when resolution adaptation is implemented (Section~\ref{sec:adapt_res}).
Finally, we give concluding remarks in Section~\ref{sec:outlook}.

\section{Lightcones}
\label{sec:data}

We work with two lightcone (LC) datasets simulated with the
semi-numerical code 21cmFASTv3~\cite{Murray2020}. An LC is a discrete
3-dimensional field of 21cm brightness offset temperature fluctuations
$\delta T_b(\mathbf{x},\nu)$ over on-sky coordinates $\mathbf{x}$ and
frequency $\nu$.
For our analysis, we focus on six model and simulation parameters,
two determining the cosmology, two sensitive to EoR
astrophysics~\cite{Neutsch:2022hmv}, and two to cosmic dawn
astrophysics:
\begin{itemize}
    \setlength\itemsep{8pt}
    \item $m_\text{WDM}\in [0.3,10]\,\text{keV}$: the lower limit on the
    warm dark matter mass allows for a small tension with cold dark
    matter (CDM), current astrophysical constraints point towards the
    upper limit~\cite{Villasenor:2023,Irsic:2023}. Here, structure
    formation looks similar to CDM, because the free-streaming length is
    inversely proportional to $m_\text{WDM}$;
    \item $\Omega_\text{m} \in [0.2,0.4]$: the dark matter density
    parameter controls structure formation. The range for training is
    deliberately chosen wider than Planck limits~\cite{Planck:2018vyg}
    but encloses them;
    \item $E_0 \in [100,1500]\,\text{eV}$: the X-ray energy threshold for
    self-absorption by host galaxies, where X-rays with energies below
    $E_0$ do not escape the host galaxy;
    \item $L_\text{X} \in
    [10^{38},10^{42}]\,\text{erg}\,\text{s}^{-1}\,\text{M}_\odot^{-1}
    \,\text{yr}$: the specific integrated X-ray luminosity
    $<2\,\text{keV}$ per unit star formation rate that escapes host
    galaxies;
    \item $T_\text{vir} \in [10^4,10^{5.3}]\,\text{K}$: the minimum virial
    temperature (related to a minimal virial halo mass) needed for
    cooling within halos to enable star formation;
    \item $\zeta\in [10,250]$: the ionization efficiency, given by 
    \begin{align}
        \zeta = 30\,\frac{f_\text{esc}}{0.3} \; \frac{f_\star}{0.05} \; 
        \frac{N_{\gamma/b}}{4000} \; \frac{2}{1+n_\text{rec}} \; ,
    \end{align}
    in terms of the escape fraction of ionizing photons into
    the intergalactic medium $f_\text{esc}$, the fraction of galactic
    gas in stars $f_\star$, the number of ionizing photons per baryon
    in stars $N_{\gamma/b}$, and the typical number density of
    recombinations for hydrogen in the intergalactic medium
    $n_\text{rec}$.
\end{itemize}
Our simulations sample parameters points from flat priors in the ranges given above. For all other cosmological parameters we refer to the Planck
measurements, assuming flatness and a cosmological constant. The
central values are $\Omega_\text{b}=0.04897$, $\sigma_8 = 0.8102$,
$h=0.6766$, and $n_s=0.9665$~\cite{Planck2018}.

\subsection{Datasets}
\label{sec:data_res}

\paragraph{High-resolution (HR)}
The first of our datasets~\cite{Neutsch:2022hmv,Schosser:2024aic}
consists of 5k LCs with spatial size $200\times200~\text{Mpc}^2$ and
redshift range $z\in[5, 35]$. It is simulated at a spatial resolution
of 1.42~Mpc, leading to LCs with 140 voxels along each on-sky axis and
variable length (depending on \Om) in the redshift axis. To
standardize the LC shapes, we keep the first 2350 voxels.
Consequently, only LCs with $\Om=0.4$ entirely span $z\in[5,35]$,
while the rest end at $z<35$. Since the prior ranges given above are
conservative, a small fraction of LCs display unrealistic reionization
histories. For example, some parameter combinations result in LCs with
a Thomson scattering optical depth inconsistent with
Planck~\cite{Planck2018} at more than $5\sigma$, or late reionization
such that the mean intergalactic-medium neutral fraction
$\bar{x}_\mathrm{HI}$ is over 0.1 at redshift $z\sim5$, in strong tension with
Ly$\alpha$ forest observations~\cite{Qin:2021}. Our 5k LCs are
filtered to satisfy these criteria.

\paragraph{Noised LCs}
Mock observed noised realizations of the 5k HR LCs are created by
splitting each LC into smaller coeval boxes than were used during simulation. Each coeval box corresponds to a fixed redshift. At each
redshift the expected noise power is estimated using
21cmSense~\cite{Pober_2013,Pober_2014}. The thermal and instrumental noise estimate is
based on $\sim$1000 hrs of SKA-Low stage 1 tracked observations, and the noise power is added to the Fourier-transformed coeval boxes. We follow a foreground avoidance strategy where galactic and extragalactic foregrounds to the 21cm signal are localized in $k$-space in the so-called 21cm foreground wedge~\cite{Morales:2012kf,Liu:2014yxa}; we assume the wedge covers the primary field-of-view of the instrument. These modes are zeroed out before transforming the full boxes of signal plus noise to real space to obtain noised LCs.

\begin{table}
    \centering \setlength{\tabcolsep}{12pt}
    \begin{small} \begin{tabular}{lccc}
        \toprule
        Dataset & HR & HRDS & LR\\
        \midrule
        LC Shape & $(140, 140, 2350)$ & $(28, 28, 470)$ &  $(28, 28, 470)$\\
        Simulated resolution [Mpc] & 1.42 & 1.42 & 2.84 \\
        Downsample factor & - & 5 & 2.5 \\
        Noised version available & \cmark & \cmark & \xmark \\
        Filter valid LCs & \cmark & \cmark & \xmark \\
        Total LCs & 5k & 5k & 35k\\
        \bottomrule
    \end{tabular} \end{small}
    \caption{Details of the high-resolution (HR), HR-downsampled
      (HRDS), and low-resolution (LR) lightcone datasets.}
    \label{tab:datasets}
\end{table}

\paragraph{Low-resolution (LR)}
A further 35k LCs were simulated using identical prior ranges for the
parameters, but with a coarser resolution of 2.84~Mpc. This yields lightcones with
shape $(70,70,1175)$. To alleviate computational demand, we then downsample these images by a factor of 2.5 in each axis using
{\tt transforms.resize} from the {\tt skimage} package. This results in lightcones with shape $(28,28,470)$. Unlike the HR
datasets, we do not make any selection based on the validity of the
reionization history.

\paragraph{HR-downsampled (HRDS)}
Finally, we construct a downsampled version of the HR dataset with image size $(28,28,470)$ to match the LR dataset. This is achieved by averaging over $5\times5\times5$ groups of voxels. This HRDS dataset is superior to the LR dataset, since the information available to the forward simulation degrades with lower simulated resolution. As we will see
later, the main difference is that the LR dataset has no sensitivity
to \mWDM simply because of its reduced simulation resolution. This parameter,
as well as \Tvir, place a threshold on early star formation, via the
Jeans mass for \mWDM, and are thus degenerate. Because the LR dataset
does not include information on \mWDM, the second parameter \Tvir can
be extracted perfectly without this parameter degeneracy present.
Once limited knowledge about \mWDM is introduced into the HRDS
dataset, the extraction of \Tvir becomes much less precise and prone
to outliers. For further discussion, see Appendix~\ref{app:degeneracy}.



A summary of the three datasets is given in
Table~\ref{tab:datasets}. In the LR dataset we reserve 1k LCs for
testing and 25k LCs for training, with the remainder used as
validation. For the HR and HRDS datasets, the testing splits consist of 750 LCs
and up to 3.75k are used for training.

\subsection{Preprocessing and augmentations}
\label{sec:data_proc}

We perform simple shift and scale transformations to
preprocess LC voxel values and parameter labels into the range
$[0,1]$,
\begin{align}
    x \; \to \; \frac{x-x_\text{min}}{x_\text{max}-x_\text{min}} \;.
\end{align}
For the simulation parameters, the minimum and maximum values
correspond to the prior boundaries. For the voxels we use
$(x_\text{min},x_\text{max})=(-120,1)$ universally.

To boost the statistical power of the datasets, we employ data
augmentation based on symmetries of the LCs. At each training
iteration we randomly sample a transformation by composing 90$^\circ$
rotations around the redshift axis with an optional reflection in the
spatial axes. The identity is included in the set of possible
transformations. These augmentations are beneficial for every task
discussed below.

\section{\skatr}
\label{sec:skatr}

\subsection{Vision transformer}
\label{sec:skatr_arch}

Transformers are a powerful network architecture for processing sequence data~\cite{NIPS2017_3f5ee243} and have proven useful in fundamental LHC physics~\cite{Plehn:2022ftl,Butter:2023fov}. They can be adapted to non-sequence data by using specialized positional encodings, which are necessary to break the permutation-equivariance of the attention mechanism. For example, constructing encodings based on a grid in two or more dimensions allows application to images, leading to the Vision Transformer (ViT)~\cite{dosovitskiy2021an,Favaro:2024rle}.

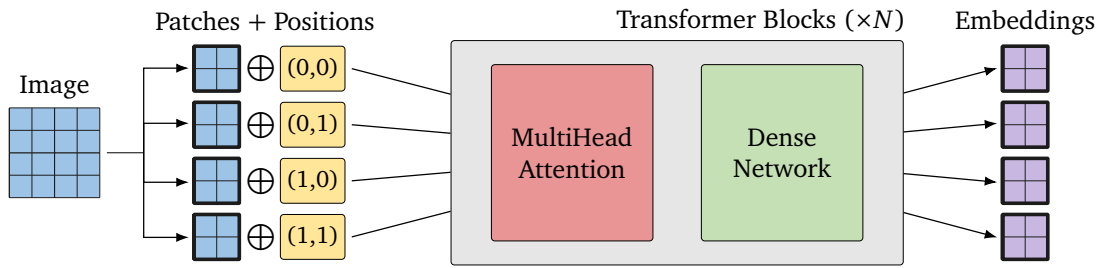
\begin{figure}[b!]
    \centering
    \def\figscale{0.4}
\def\picscale{0.35}
\def\linecolor{gray!178}
\def\ctxcolor{Gcolor}
\def\tgtcolor{Rcolor}

\scalebox{0.85}{\begin{tikzpicture}[scale=\figscale, font=\large]
    
    \tikzset{
    pics/patch/.style args={#1/#2}{
      code = {
        \begin{scope}[scale=\picscale, shift=({-1, -1})]
            \draw[\linecolor, #2, rounded corners=0.3pt, fill=#1, fill opacity=1.]
            (0,0) -- (0,2) -- (2,2) -- (2,0) -- (0,0) -- (0,2);
            \draw[\linecolor] (0,1) -- (2,1);
            \draw[\linecolor] (1,0) -- (1,2);            
        \end{scope}
    }}}

    \tikzset{
    image/.pic={
        \begin{scope}[scale=\picscale]
            \foreach \x in {-1, 1} {
                \foreach \y in {-1, 1} {
                    \draw (\x,\y) pic {patch=Bcolor/thin};
                }
            }
        \end{scope}
    }}

    \tikzset{
    pics/patchset/.style args={#1}{
      code = {
        \begin{scope}[scale=\picscale]
            \foreach \y in {-3.75, -1.25, 1.25, 3.75} {
                \draw (0,\y) pic {patch=#1/ultra thick};
            }
        \end{scope}
    }}}


    \node (block) [expr, minimum width=7cm, minimum height=3.5cm, fill=gray!20, label={[xshift=1.3cm]Transformer Blocks ($\times N$)}] at (0,0) {};

    \node (image) [left=6cm of block, label={[yshift=0.575cm]Image}] {};
    \draw (image) pic {image};

    \node (inpatches) [right=2.25cm of image, label={[yshift=1.655cm, xshift=0.75cm]Patches + Positions}] {};
    \draw (inpatches) pic {patchset=Bcolor};    

    \node (p1) [expr, minimum height=0.7cm, minimum width=1cm, outer sep=3pt, right=0.75cm of inpatches, yshift=1.3125cm, font=\normalsize, label={[yshift=-0.775cm, xshift=-0.8cm]$\bigoplus$}] {(0,0)};
    \node (p2) [expr, minimum height=0.7cm, minimum width=1cm, outer sep=3pt, right=0.75cm of inpatches, yshift=0.4375cm, font=\normalsize, label={[yshift=-0.775cm, xshift=-0.8cm]$\bigoplus$}] {(0,1)};
    \node (p3) [expr, minimum height=0.7cm, minimum width=1cm, outer sep=3pt, right=0.75cm of inpatches, yshift=-0.4375cm, font=\normalsize, label={[yshift=-0.775cm, xshift=-0.8cm]$\bigoplus$}] {(1,0)};
    \node (p4) [expr, minimum height=0.7cm, minimum width=1cm, outer sep=3pt, right=0.75cm of inpatches, yshift=-1.3125cm, font=\normalsize, label={[yshift=-0.775cm, xshift=-0.8cm]$\bigoplus$}] {(1,1)};

    \node (attn) [expr, minimum width=2.5cm, minimum height=2.75cm, fill=Rcolor, xshift=-1.625cm] at (block){MultiHead\\Attention};
    \node (dense) [expr, minimum width=2.5cm, minimum height=2.75cm, fill=Gcolor, xshift=1.625cm] at (block){Dense\\Network};

    \node (outpatches) [right=1.75cm of block, label={[yshift=1.575cm,align=center]Embeddings}] {};
    \draw (outpatches) pic {patchset=Pcolor};

    \node (node) [black, right=1.25cm of image, inner sep=0, outer sep=0, minimum size=0] {};
    \draw ([xshift=1.7cm]image.east) -- (node);
    \draw[arrow] (node) |- ([xshift=-0.75cm, yshift=3.3cm]inpatches.west);
    \draw[arrow] (node) |- ([xshift=-0.75cm, yshift=1.15cm]inpatches.west);
    \draw[arrow] (node) |- ([xshift=-0.75cm, yshift=-1.15cm]inpatches.west);
    \draw[arrow] (node) |- ([xshift=-0.75cm, yshift=-3.3cm]inpatches.west);

    \draw[arrow] (p1.east) -- (block) -- ([xshift=-0.75cm, yshift=3.3cm]outpatches.west);
    \draw[arrow] (p2.east) -- (block) -- ([xshift=-0.75cm, yshift=1.15cm]outpatches.west);
    \draw[arrow] (p3.east) -- (block) -- ([xshift=-0.75cm, yshift=-1.15cm]outpatches.west);
    \draw[arrow] (p4.east) -- (block) -- ([xshift=-0.75cm, yshift=-3.3cm]outpatches.west);
    
\end{tikzpicture}}
    \caption{Schematic diagram of a vision transformer encoder using $2\times2$ patches.}
    \label{fig:vit-flow}
\end{figure}

Figure~\ref{fig:vit-flow} shows how a ViT processes images. First, an image is divided into non-overlapping patches of pixels. Each patch is embedded
into a high-dimensional space using a shared linear projection, then
augmented with an encoding of the patch location in the image. The
set of patch representations are processed by an alternating sequence
of multi-head attention and feed-forward operations. Normalization layers and skip connections are also used to stabilize optimization. 

In a ViT, the patch size controls a trade-off between model
expressivity and complexity. While using smaller patches probes
spatial correlations in the input image at finer scales, this also
leads to a larger number of elements entering the attention
operation. In order to manage the computational cost of a ViT, the
patch size should be selected with an expected image resolution in
mind. In our analysis, we use patch sizes of (4,4,10) for downsampled
LCs, and (7,7,50) for full-resolution LCs in the HR dataset. We arrive at these values by hand; they approximately maximize the total number of patches (with a similar number in each axis) while remaining within memory limits.

Depending on the task, a loss can be calculated directly using the set
of patch embeddings output by the ViT. This is the case for our pretraining, to be presented in the following section, as well as other common tasks such as segmentation or diffusion. For cases where a global feature vector is needed, such as in regression, an aggregation step can be used to obtain a single input for the task-specific head network. There are a number of possibilities for this aggregation. A simple option is a mean over patch embeddings $z$ followed by a
two-layer dense network,
\begin{align}
    \text{MLP}(z) \equiv W_2 \ReLU (W_1 \langle z\rangle)\;,
    \label{eq:mlp_head}
\end{align}
where $\langle z\rangle$ is the average of $z$ over patches, and $W_1$ and $W_2$ are weight matrices with respective shapes $d\times d$ and $6\times d$ for transformer embedding dimension $d$. We use this setting for all regression results in our analysis, where the output dimension 6 corresponds to the number of target parameters. 
Another, more flexible, possibility is to learn a dynamic pooling function using a cross
attention layer,
\begin{align}
    \operatorname{XAttn}(z) \equiv \operatorname{Softmax}\left(\frac{q^TW_Kz}{\sqrt{d}}\right)W_Vz\;,
    \label{eq:xattn_head}
\end{align}
with $d\times d$ key and value weight matrices $W_K$ and $W_V$, and a learnable $d\times1$ input token $q$. For the results in this work, we find the simple mean and MLP option to be sufficient.

\subsection{Self-supervised pre-training}
\label{sec:skatr_self}

Lightcones are represented in a high-dimensional voxel space, which
we expect to be compressible. For any kind of analysis, we need
encoders to map the voxels to a lower-dimensional embedding space.
The goal of self-supervised pre-training is to learn an encoding~
$f_\theta(x)$ that produces informative representations of the data
$x$ without using any labels or, in our case, model
parameters. 

\begin{figure}[t!]
    \centering
    \def\figscale{0.4}
\def\picscale{0.35}
\def\linecolor{gray!178}
\def\ctxcolor{Gcolor}
\def\tgtcolor{Rcolor}

\scalebox{0.85}{\begin{tikzpicture}[scale=\figscale, font=\large]
    
    \tikzset{
    pics/patch/.style args={#1}{
      code = {
        \begin{scope}[scale=\picscale]
            \draw[\linecolor, ultra thick, rounded corners=0.3pt, fill=#1, fill opacity=1.]
            (0,0) -- (0,2) -- (2,2) -- (2,0) -- (0,0) -- (0,2);
            \draw[\linecolor, ultra thin] (0,1) -- (2,1);
            \draw[\linecolor, ultra thin] (1,0) -- (1,2);            
        \end{scope}
    }}}

    \tikzset{
    image/.pic={
        \begin{scope}[scale=\picscale, shift=({-3, -3})]
            \foreach \x in {0, 2, 4} {
                \foreach \y in {0, 2, 4} {
                    \def\patchfill{\ctxcolor}
                    \ifthenelse{\x=4}{\def\patchfill{\tgtcolor}}
                    {\ifthenelse{\x=2 \AND \y=4}{\def\patchfill{\tgtcolor}}}
                
                    \draw (\x,\y) pic {patch=\patchfill};
                }
            }
        \end{scope}
    }}  

    \tikzset{
    context/.pic={
        \begin{scope}[scale=\picscale, shift=({-2, -3})]
            \foreach \x in {0, 2, 4} {
                \foreach \y in {0, 2, 4} {
                    \ifthenelse{\x=4}{}{
                    \ifthenelse{\x=2 \AND \y=4}{}{
                    \draw (\x,\y) pic {patch=\ctxcolor};
                    }}                
                }
            }        
        \end{scope}
    }} 
    
    \tikzset{
    target/.pic={
        \begin{scope}[scale=\picscale, shift=({-4, -3})]
            \foreach \x in {0, 2, 4} {
                \foreach \y in {0, 2, 4} {
                    \ifthenelse{\x=4}{
                        \draw (\x,\y) pic {patch=\tgtcolor};
                    }{
                    \ifthenelse{\x=2 \AND \y=4}{
                        \draw (\x,\y) pic {patch=\tgtcolor};
                    }}                
                }
            }    
        \end{scope}
    }}  

    \node (image) [label={[yshift=1cm]Input $x$}] at (0,0) {};
    \draw (image) pic {image};
    \node (context) [below=3.5cm of image, label={[yshift=-1.9cm] $\Tilde x$}] {};
    \draw (context) pic {context};
    
    \node (ctx_b) [net_black, right=3cm of context] {};
    \node (ctx_net) [net] at (ctx_b) {\normalsize{Context}\\\normalsize{encoder} \\ $f_\theta$ };

    \node (tgt_b) [net_black, right=3cm of image] {};
    \node (tgt_net) [net] at (tgt_b) {\normalsize{Target}\\\normalsize{encoder}\\ $g_\varphi$};

    \node (ctx_out) [right=3cm of ctx_b, label={[yshift=-1.9cm] $\Tilde z$}] {};
    \draw (ctx_out) pic {context};
    \node (tgt_out) [right=3cm of tgt_b, label={[xshift=1.1cm, yshift=-0.2cm] $z$}] {};
    \draw (tgt_out) pic {target};

    \node (prd_b) [net_black, right=3cm of ctx_out, label={[yshift=-1.9cm] $\Tilde z$}] {};
    \node (prd_net) [net] at (prd_b) {\normalsize{Predictor} \\ $h_\psi$};

    \node (prd_out) [above=2.47cm of prd_net, label={[xshift=1.1cm, yshift=-0.28cm] $p$}] {};
    \draw (prd_out) pic {target};

    \node (loss) [expr, right=0.38cm of tgt_out, yshift=2.5cm, outer sep=4pt, fill=gray!30, minimum width=3cm] {\large $\loss_\text{JEPA} = |z-p|$ };

    \draw [arrow, color=\linecolor] ([yshift=-3cm]image.south)-- ([yshift=3cm]context.north);
    \draw [arrow, color=\linecolor] ([xshift=2cm]context.east) -- (ctx_net) -- ([xshift=-2cm]ctx_out.west);
    \draw [arrow, color=\linecolor] ([xshift=3cm]image.east) -- (tgt_net) -- ([xshift=-2cm]tgt_out.west);
    \draw [arrow, color=\linecolor] ([xshift=2cm]ctx_out.east) -- (prd_net) -- ([yshift=-3cm]prd_out.south);

    \draw [arrow, color=\linecolor] ([yshift=3cm]tgt_out.north) |- (loss);
    \draw [arrow, color=\linecolor] ([yshift=3cm]prd_out.north) |- (loss);
    
\end{tikzpicture}}
    \caption{Illustration of the pre-training. At each training
      iteration, a mask is sampled that defines context (green) and
      target (red) patches. After training, the context encoder is
      taken as the summary network.}
    \label{fig:jepa}
\end{figure}
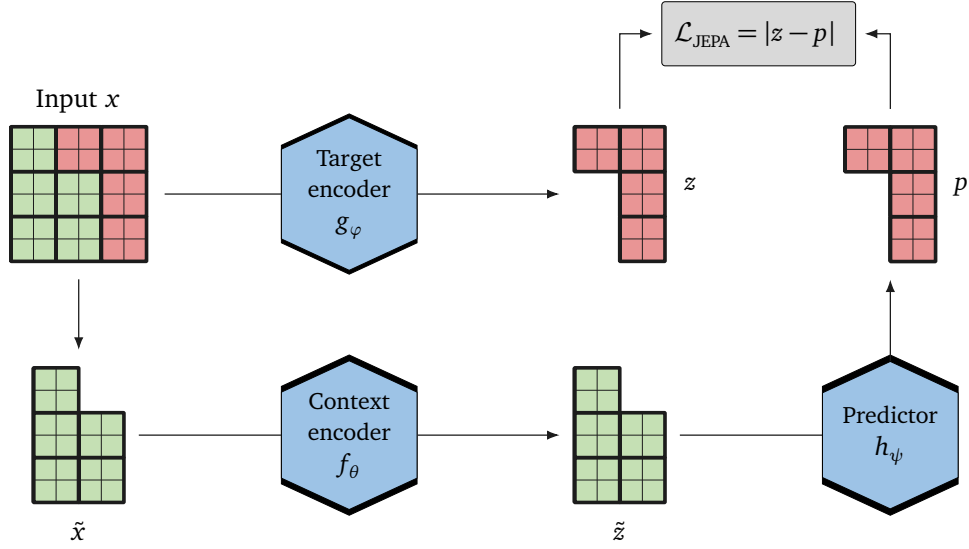

To train such an encoder, we adopt a self-supervised learning
framework based on Joint Embedding Predictive Architectures
(JEPA)~\cite{Assran_2023_CVPR, bardes2024revisiting}. Our
SKA Transformer (\skatr) setup is shown in Figure~\ref{fig:jepa} and
involves two ViTs: a context encoder $f_\theta$ and a target encoder
$g_\varphi$. The networks share identical architectures and are
initialized with the same weights. During training, an LC
(batch) is divided into a set of $n$ patches $x=\{x_i\}_{i=1}^n$. A
masked view~$\Tilde x$ is generated by dropping a sampled set of patch
locations $M$
\begin{align}
    \label{eq:masking}
    \Tilde x = \{x_i \in x \,|\, i\notin M\} \subset x\; .
\end{align}
The masked and original LCs are embedded with the context and
target networks respectively,
\begin{align}
  \Tilde z = f_\theta(\Tilde x)
  \qquad \text{and} \qquad
  z = g_\varphi(x)\; .
\end{align}
Finally, a transformer $h_\psi$ with smaller hidden dimension than the encoders predicts the target patch
embeddings given the context embeddings,
\begin{align}
    p = h_\psi(\Tilde z)\; .
\end{align}
The loss is the mean absolute error between $p$ and $z$ at
the locations of the masked patches,
\begin{align}
    \label{eq:pretrain_loss}
    \loss_\text{JEPA} = \left\langle\frac{1}{|M|}\sum_{i\in M}\Big|g^i_\varphi(x) - h^i_\psi(f_\theta(\Tilde x)) \Big| \right\rangle_{p_\text{data}(x),\,p_\text{mask}(M)}\,,
\end{align}
where $p_\text{mask}(M)$ encodes a user-defined masking strategy (see
Appendix~\ref{app:training-details} for details). Given that the above loss is not contrastive, there is a risk of representation collapse if we
optimize it with respect to all parameters $\theta, \varphi$, and $\psi$. To
avoid this, only the context encoder and predictor parameters are
updated via gradient descent. The target encoder parameters instead
follow the exponential moving average of the context encoder
parameters,
\begin{align}
    \varphi_{i+1} = \tau \varphi_i + (1-\tau)\,\theta_i\; , 
\end{align}
where $\tau$ is a momentum hyperparameter controlling the rate at
which the target encoder parameters are updated.

Once trained, a global summary of an LC can be constructed by
passing it through the context encoder without a mask, then taking the
mean over the resulting patch embeddings. Throughout our analysis we
always pre-train \skatr on the LR dataset, with an embedding dimension of 360. This leads to a highly-compressed \skatr representation in $\sim10^3$ times fewer dimensions than the original LC. We summarize the proposed training pipeline in Figure~\ref{fig:skatr-summary}. For the complete set of hyperparameters, see Appendix~\ref{app:training-details}.

\begin{figure}
    \centering
    \includegraphics[width=\textwidth]{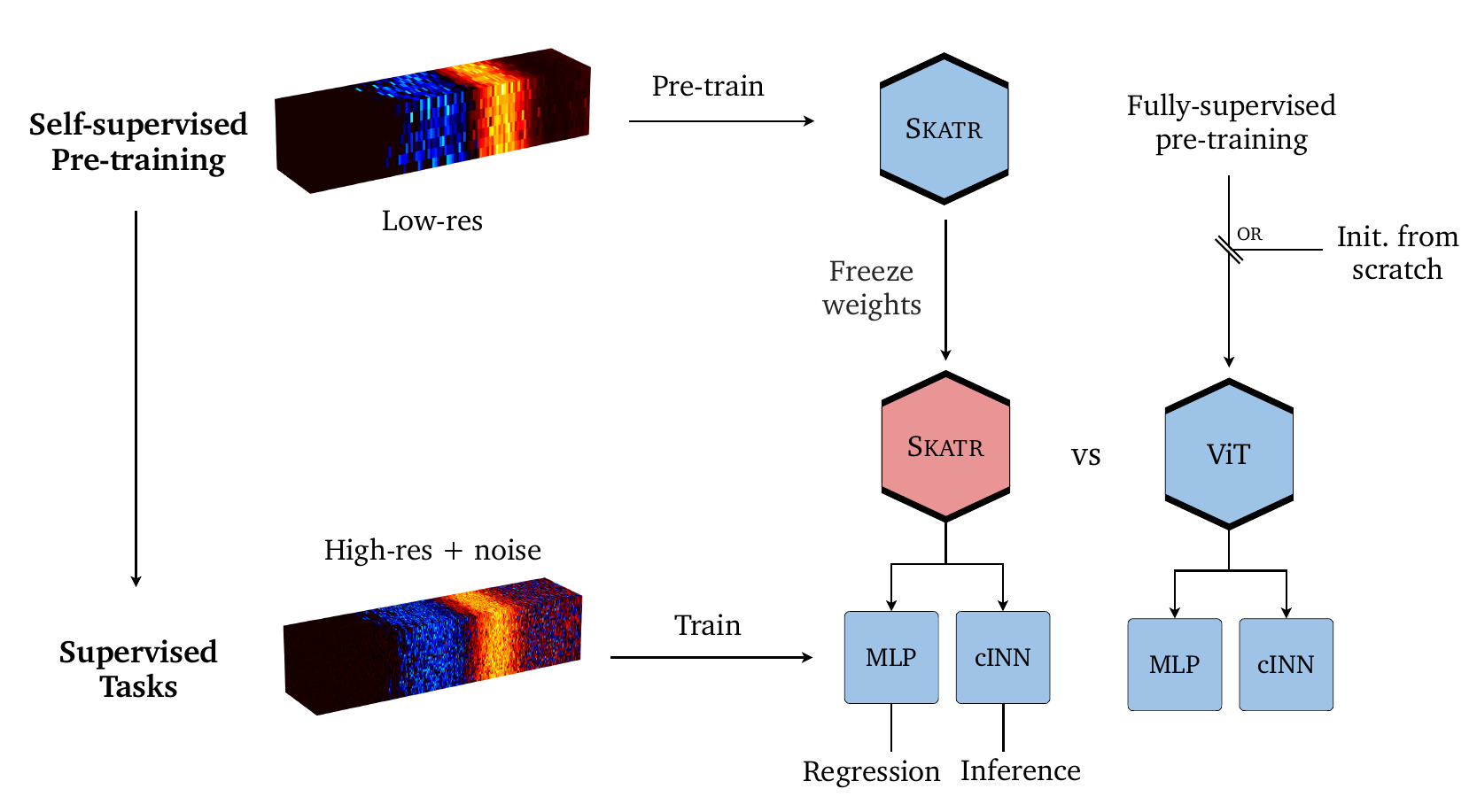}
    \caption{Summary of the pipeline for our self-supervised SKA Transformer (\skatr). Red shading indicates that the \skatr backbone is frozen for supervised tasks.}
    \label{fig:skatr-summary}
\end{figure}

\section{Results}
\label{sec:res}

To understand and quantify the behavior of \skatr as an SKA summary network we use regression and inference of six cosmological and astrophysical
parameters, described in some detail in Section~\ref{sec:data},
\begin{align}
  y\equiv\left\{ \; 
   \mWDM, \Om, E_0, \LX, \Tvir, \zeta
   \; \right\} \; .
\end{align}
The benchmark for this regression task is given by the 3D-21cmPIE-Net, a CNN~\cite{Neutsch:2022hmv,Heneka:2023}. This network is
also used as a summary network for the corresponding generative
inference~\cite{Schosser:2024aic}, where it serves the same function
with respect to data compression as a foundation model. In all cases below, the \skatr network is pre-trained as described in Section~\ref{sec:skatr}, then frozen.

\subsection{ViT performance}
\label{sec:res_vit}

\begin{figure}[t]
    \includegraphics[width=\textwidth]{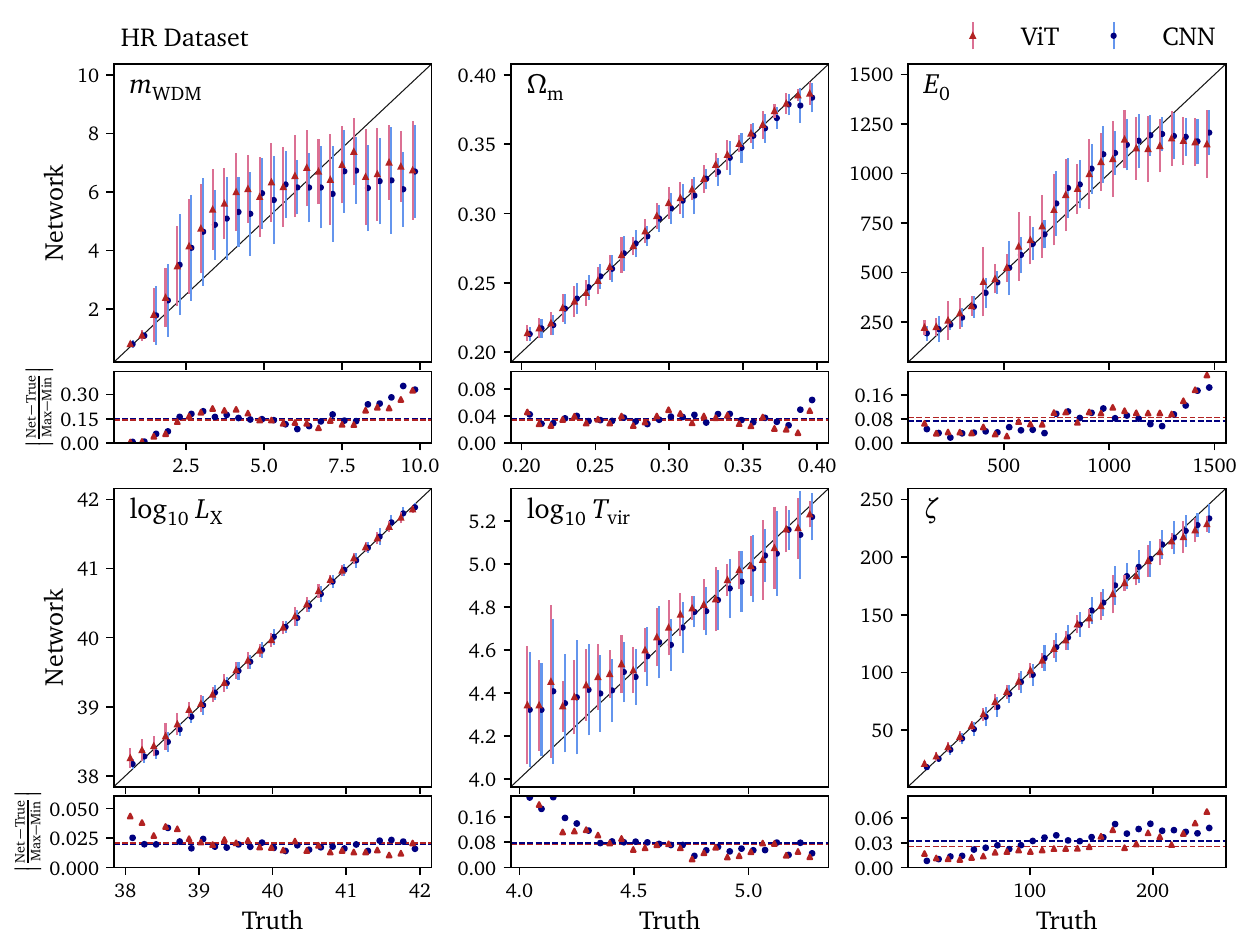}
    \caption{Performance for our \textbf{ViT (red) vs
        the CNN benchmark (blue)}, both trained from scratch to regress simulation parameters on the HR
      dataset. Network predictions on the test set are binned by the
      true parameter value, and points show the mean$\;\pm\;1\sigma$
      in each bin. The sub-panel shows the mean absolute error per bin
      normalized to the simulated parameter ranges.}
    \label{fig:cnn-vit-reg-hires}
\end{figure}

First, we demonstrate the power of the ViT by comparing with the
established CNN~\cite{Neutsch:2022hmv,Heneka:2023} for the HR
dataset. The CNN consists of a series of 3D-convolutional blocks that gradually downsample input LCs, followed by global mean and a 4-layer MLP. We train both networks to regress our six simulation
parameters using the normalized mean absolute error (NMAE) loss,
\begin{align}
    \text{NMAE} = \Bigg|\frac{y_\text{pred}-y_\text{true}}{y_\text{max} - y_\text{min}}\Bigg|\;.
\end{align}
We also measure performance in terms of the NMAE. This allows us to
compare errors across different parameters, irrespective of their
magnitude or simulated prior range.

The regression results using the HR dataset are shown in
Figure~\ref{fig:cnn-vit-reg-hires}, where the predictions of both
networks are shown against the true value for the simulated
lightcone in each parameter. To improve visual clarity, points are
binned by the true parameter label and the CNN and ViT predictions are
slightly offset from one another horizontally. The lower subpanels show the NMAE following the same binning (without error bars), and the horizontal dashed line indicates the mean error over all test points.

The ViT and the CNN benchmark both regress \Om, \LX and $\zeta$ well, while the remaining
parameters are more difficult. In particular, the LCs lose sensitivity
to \mWDM and $E_0$ above thresholds of 3~keV and 1~keV
respectively. At these points the network predictions plateau. \Tvir is
regressed poorly due to its degeneracy with \mWDM, as discussed in
Appendix~\ref{app:degeneracy}. Comparing the two networks, we see that, the
ViT indeed extracts information from the LCs
at a level at least as strong as the 21cmPIE-Net.

\subsection{\skatr regression}
\label{sec:res_skatr}

Now we examine the performance of the LC summary learned by
\skatr. We again look at parameter regression, but this time comparing
a ViT (trained from scratch) to a lightweight network trained on
\skatr-summarized LCs using the LR dataset. Since the \skatr backbone
will not be trained in this stage, we evaluate the summary once on
each LC and save the resulting dataset. This dataset is then
used to train the small network, whose architecture we match with the
2-layer MLP from Eq.~\eqref{eq:mlp_head}. To implement data augmentation
in this scheme, we also summarize all transformations of a given
lightcone (see Section~\ref{sec:data_proc}). The data loading is then customized to select a random
transformation in each batch during training.

\begin{figure}
    \includegraphics[width=\textwidth]{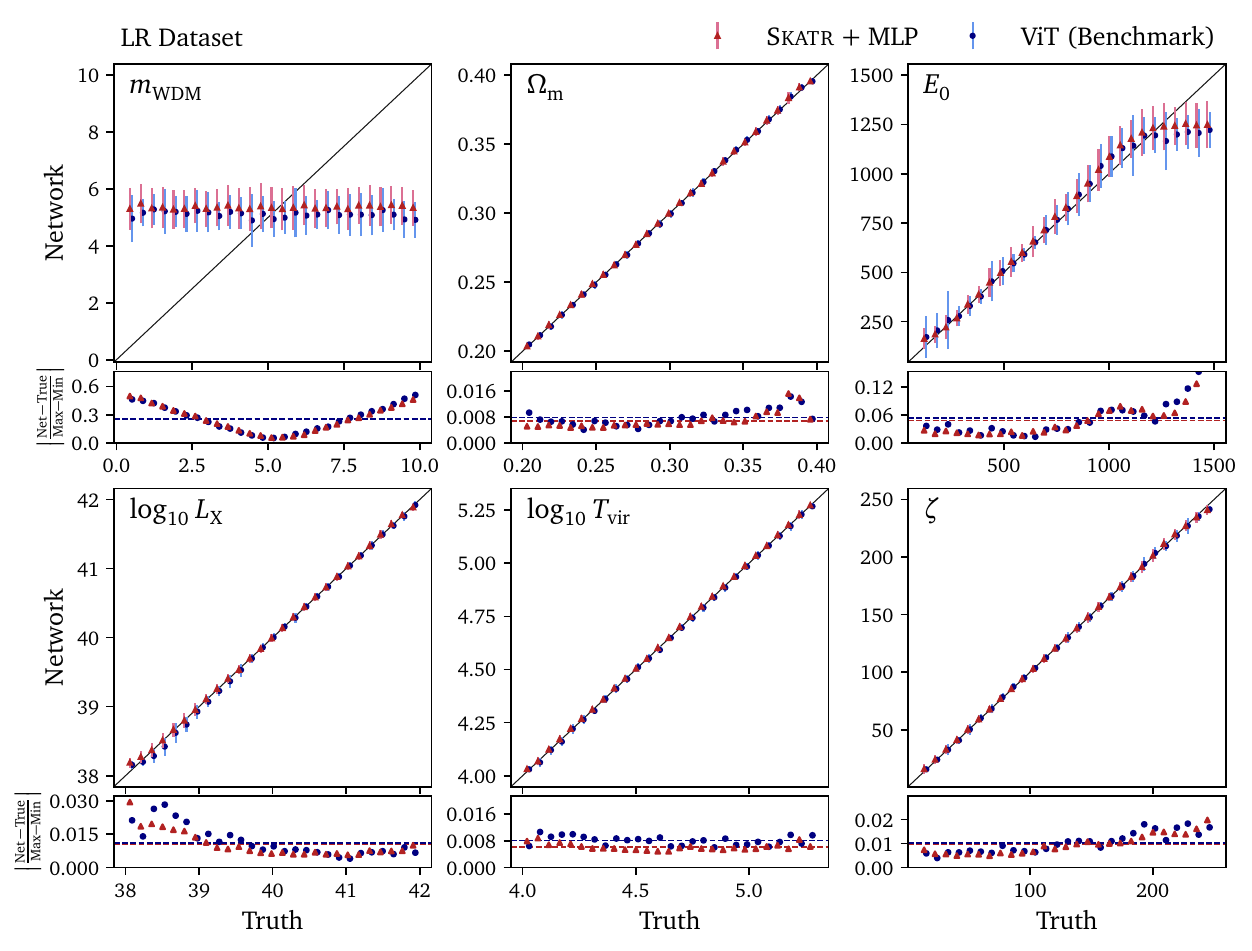}
    \caption{Performance for 
      \textbf{frozen \skatr summaries (red) vs ViT benchmark trained from
        scratch (blue)}, where \skatr is complemented with a 2-layer MLP. All training and testing is performed on the LR dataset.}
    \label{fig:scratch-pretrained-reg}
\end{figure}

The results for the LR dataset are shown in
Figure~\ref{fig:scratch-pretrained-reg}. Due to the coarse resolution,
\mWDM is no longer predictable at any point in the prior range. This
resolves the correlations and degeneracies present in the HR dataset
and allows both networks to regress the \Om, \LX, \Tvir, and $\zeta$
parameters extremely precisely and slightly improve $E_0$.

\begin{figure}
    \centering
    \includegraphics[width=0.7\linewidth]{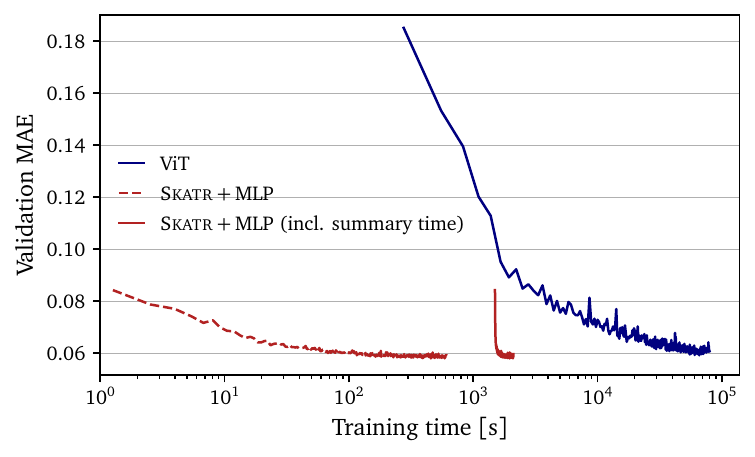}
    \caption{Loss over time for an
    \textbf{MLP with frozen \skatr summaries (red) vs ViT trained from
    scratch (blue)}, as shown in Figure~\ref{fig:scratch-pretrained-reg}. 
    The solid \skatr line includes the time to summarize the dataset, while the dashed line does not. Each line begins at the end of the first epoch.}
    \label{fig:finetune_timing}
\end{figure}

Comparing the ViT, trained from scratch, with the MLP acting on
\skatr-summarized LCs, the compressed representation achieves equal or
smaller average error. This
demonstrates that the \skatr summary retains all relevant information.
Further, because the MLP has so few parameters, its training is more stable and converges much faster than the ViT.
In Figure~\ref{fig:finetune_timing} we illustrate this acceleration using the
validation loss. For training alone, \skatr leads to a speed enhancement 
by a factor of several hundreds, to reach the same performance as the 
ViT. Even including the time to summarize the dataset, \skatr is roughly 
a factor 50 faster in training to the final converged performance.

\subsection{\skatr inference}
\label{sec:res_inf}

Next we test \skatr on a more challenging task --- inference of the full 6-dimensional posterior distribution $p(y|x)$ of parameters $y$ given data $x$. As in Ref.~\cite{Schosser:2024aic}, we train conditional invertible neural networks (cINNs) to approximate the posterior. The conditioning on LCs is always via a summary and so the loss is
\begin{align}
    \mathcal{L}_\text{cINN} = -\left\langle\log q_\vartheta\big(y|S_\phi(x)\big)\right\rangle_{p_\text{data}(x,y)}\,,
\end{align}
where $q_\vartheta$ is the probability density defined by the cINN, and $S$ is a summary which may be fixed or trainable, with parameters $\phi$.

For $S$, we consider two options. First, a ViT initialized from scratch with no head
network. In that case, the summary is the mean of learned patch
embeddings. Second, we use the mean embedding from a frozen
pre-trained \skatr{} network. Again, no MLP head network is used and
the cINN is conditioned directly on the frozen summary.

{
\begin{figure}
    \centering
    \includegraphics[width=0.98\textwidth]{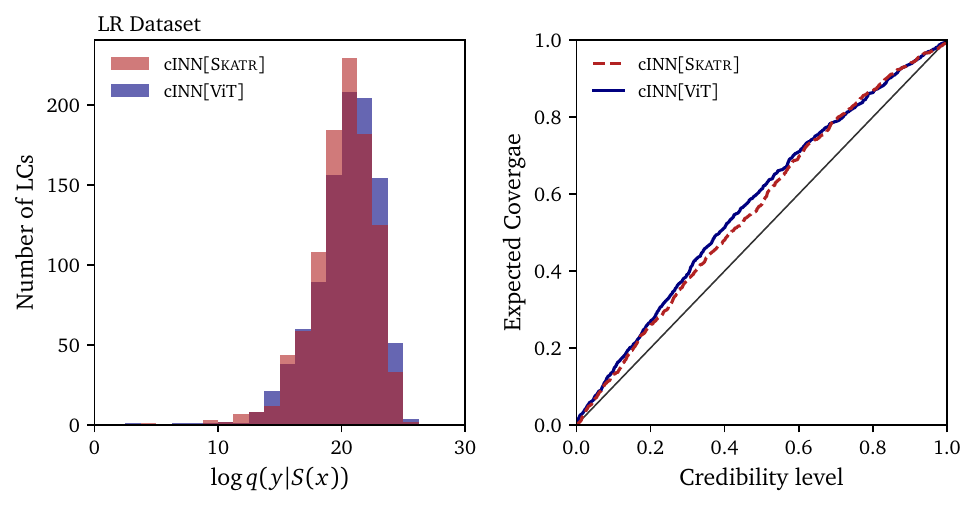}
    \caption{Inference performance for \textbf{cINN with frozen \skatr summary vs jointly-trained ViT} on the LR dataset. Shown are the distribution of log-likelihoods (left) and calibration (right) in the test set.}
    \label{fig:likelihoods_calibration_skatr}
\end{figure}%
\begin{figure}[t!]
    \vfill
    \includegraphics[width=\textwidth]{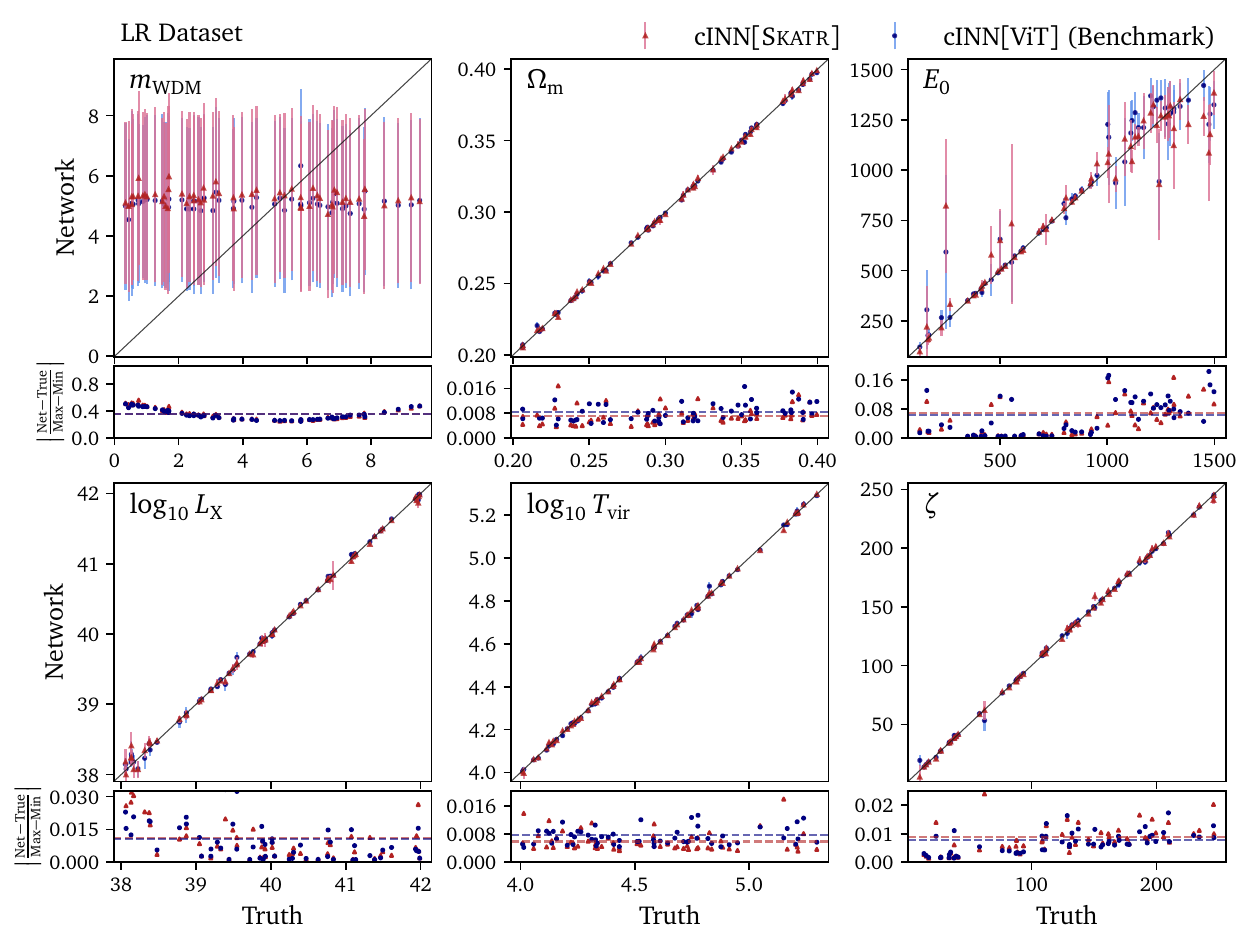}
    \caption{Posteriors using \textbf{a frozen \skatr summary vs the jointly-trained ViT benchmark}, acting as summaries for a
    cINN trained and tested on the LR dataset. For a given test LC, each point shows the posterior mean with 1$\sigma$ error bars over 5k network samples.}
    \label{fig:posterior_1d_scratch_vs_pretrain}
\end{figure}
}

To evaluate the constraining power and calibration of the networks, we
show the posterior likelihoods and coverage over the LR test set in
Figure~\ref{fig:likelihoods_calibration_skatr}. The distributions of
likelihoods (left panel) are closely matched, though a narrow advantage is evident when training from scratch with a ViT. The calibration panel (right) shows the cumulative distribution function of the rank statistic, defined as the fraction of posterior samples with model likelihood larger than the true label. In these terms, an under(over)-confident posterior lies above (below) the diagonal. We see that the posteriors defined by each network are equally conservative, i.e. slightly less precise parameter estimates than possible. In Figure~\ref{fig:posterior_1d_scratch_vs_pretrain} we show the
parameter-wise posterior predictions for a sample of LR test points,
comparing the cINN with a ViT or the frozen \skatr summary. Both networks
perform comparably to the previous results for regression. On average,
the \skatr summary gives posterior means that are closer to the truth
value in each parameter than the ViT, meaning \skatr performs at higher accuracy.

\subsection{Generalization}
\label{sec:res_gen}

Next, we demonstrate that the performance of \skatr also extends to 
new datasets. In particular, we are interested in whether \skatr
provides a benefit on LCs simulated at higher resolution than the
pre-training set, and including those with a noise model. That would allow us to transfer information from cheap
simulations to expensive simulations. To this end, we repeat our regression test with the noised HRDS dataset. Here there main challenge and interest
will be whether the shallow MLP can regress the \mWDM parameter given the \skatr summary,
as \mWDM cannot be predicted from the LR dataset used to train \skatr due to insufficient resolution.

\begin{figure}[t!]
    \includegraphics[width=\textwidth]{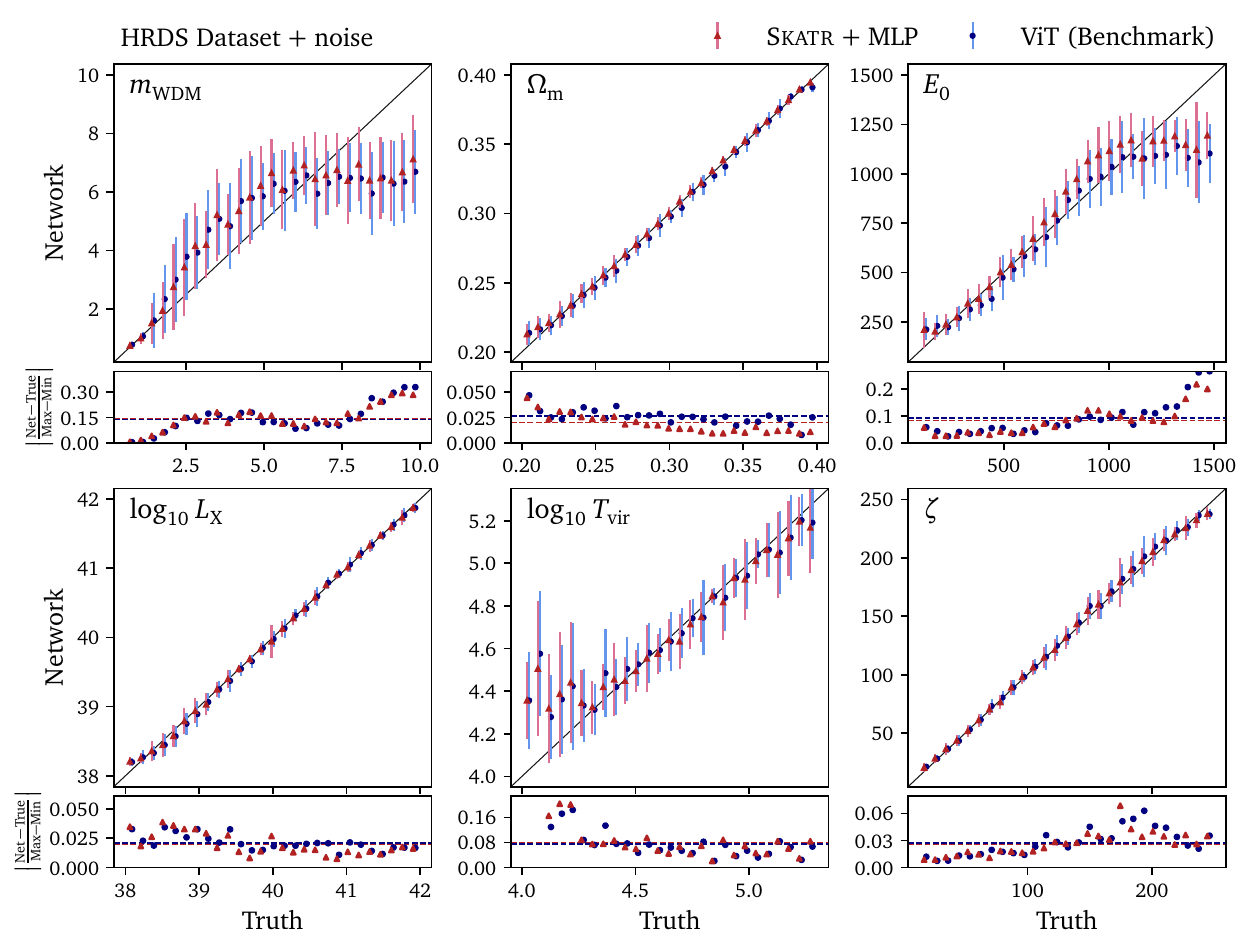}
    \caption{Performance for 
      \textbf{frozen \skatr summary vs ViT benchmark trained from
        scratch} on the noised HRDS dataset, \skatr is complemented with a trained 2-layer MLP. The \skatr pre-training on LR
      dataset does not contain any information on \mWDM.}
    \label{fig:domain-shift}
\end{figure}

In Figure~\ref{fig:domain-shift} we compare
the \skatr-MLP combination to a ViT trained from scratch on the noised HRDS dataset. Once again, we see that the \skatr summary matches the
regression performance of a full ViT training. A slight advantage for
the MLP is apparent in \Om, likely due to the large dataset used for
\skatr pre-training. Since \skatr displays no obvious failures, we conclude that the fixed summary is sufficiently general to capture new effects in the LCs due to noise and new correlations in the parameters due to simulation resolution. In particular, the \skatr-MLP combination has no
trouble regressing \mWDM, demonstrating that \skatr remains
informative outside the training domain.

\subsection{Data efficiency}
\label{sec:res_eff}

\begin{figure}[t!]
    \includegraphics[width=\textwidth]{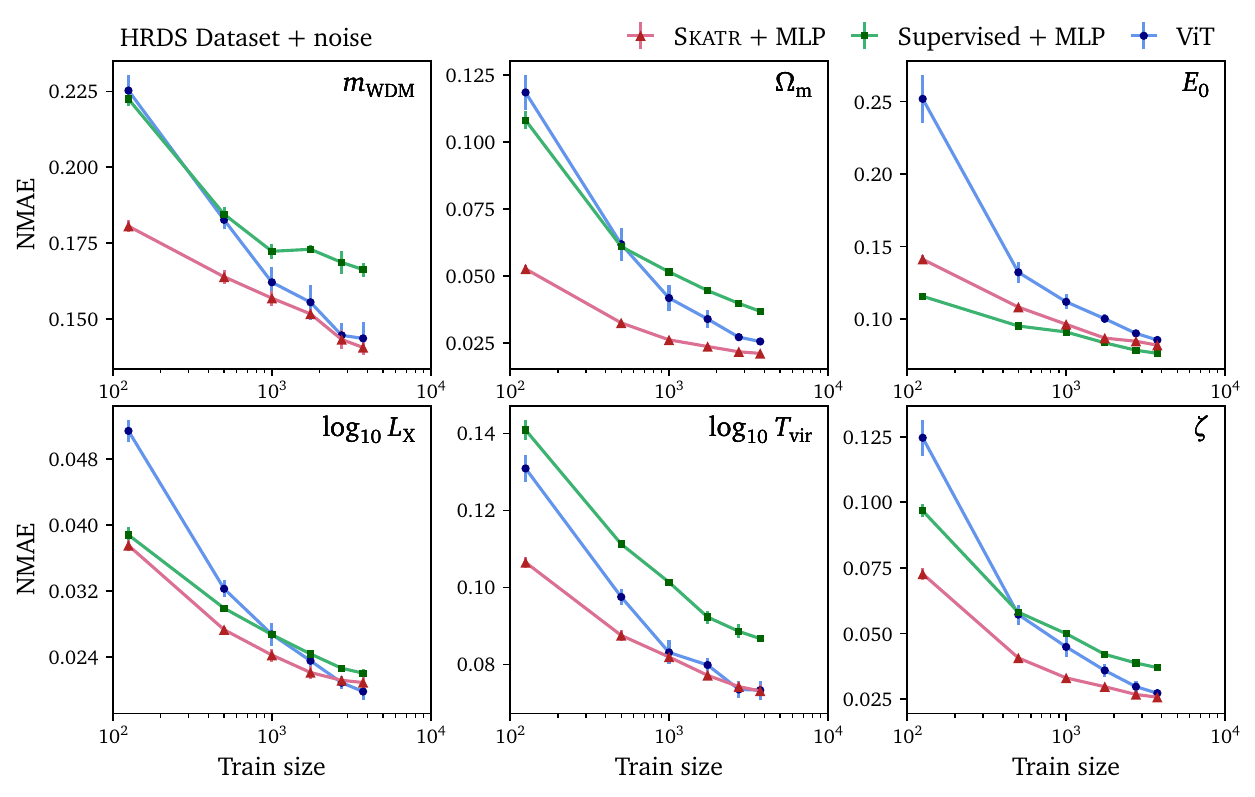}
    \caption{Data scaling for \textbf{\skatr-summarized LCs vs supervised-summarized LCs vs a ViT trained from scratch} using the noised HRDS dataset. Shown is the mean regression error as a function of the number of LCs
      used for training. Each point is the average over ten
      training runs, with $1\sigma$ error bars.
      }
    \label{fig:data-eff-hr-aug}
\end{figure}

As mentioned above, one of the main reasons for pre-training \skatr is
that it is extremely efficient when it comes to training for the
downstream task, in our case the regression of the model parameters
from some test dataset. To illustrate that gain, we emulate
data-limited scenarios by scanning a range of training split sizes
within the HRDS dataset. In each case, we train a ViT from scratch and
compare its performance to a shallow MLP trained on \skatr-compressed
LCs.  Figure~\ref{fig:data-eff-hr-aug} shows that the \skatr summary
consistently yields smaller error than the ViT trained from
scratch. This is despite being pre-trained out of domain, on LR
LCs. The most impressive improvement can be seen for little training
data, where the ViT struggles to capture the relevant information. The
only exception is in \LX, where \skatr is outperformed on the largest
training set. The improvement from \skatr over the from-scratch baseline is
smallest for \Tvir, possibly due to the degeneracy with \mWDM
introduced in the HRDS dataset. This kind of improvement can be
important for SKA, because generating HRDS training data is expensive
and will eventually limit the actual data analysis.

To see the specific impact of the self-supervised JEPA training we
introduce a second baseline, pre-training a ViT with fully-supervised
regression on the LR dataset. From this ViT we drop the regression
head and take the mean over patch embeddings as a summary.  Also in
Figure~\ref{fig:data-eff-hr-aug} we see that for most parameters \skatr
is significantly better than the supervised backbone. The only
exceptions are \LX, where the improvement is only marginal, and $E_0$,
where \skatr is slightly outperformed. Moreover, the fully-supervised summary network is typically worse than the
ViT trained from scratch. While in \Om, $E_0$, \LX and $\zeta$ the MLP
does eventually achieve a lower error in the smallest training size,
this is not the case for \mWDM or \Tvir. Recalling that these
parameters exhibit the greatest change in behavior between the LR and
HR datasets, this observation highlights the difficulty for supervised
pre-training. Instead of providing a generalizing summary, it encodes
details of the correlations relevant for the specific fully-supervised
task. When these correlations are different for the test dataset, the
regression-pre-trained model does not generalize. We have also checked that the same results hold for the pure HRDS dataset without noise.

\subsection{Resolution adaptation}
\label{sec:adapt_res}

So far, we have shown that a \skatr summary pre-trained on the LR
dataset generalizes to new correlations in the HRDS dataset which has been downsampled to match the LR resolution. What
remains to be seen is whether the LR-trained \skatr can perform on the
full-resolution HR dataset. Due to the factor $5^3$ increase in the
number of voxels between these datasets, adapting the resolution comes
with computational cost, one way or another. There are a number of
options to tackle this problem, which we now discuss in turn.

The most straightforward approach is to split the HR LCs using the
same patch shape as for pre-training. This requires no further
training of the \skatr backbone, and so a summarized dataset can be
constructed. The larger number of resulting patches can be processed
by interpolating position encodings, to preserve the total LC
size. However, for our HR dataset, the attention operation between
5$^3$ times more elements is prohibitively expensive. Further, the
physical size of patches in this scheme differs from that used in
pre-training, which is likely suboptimal.

\begin{figure}[b!]
    \includegraphics[width=\textwidth]{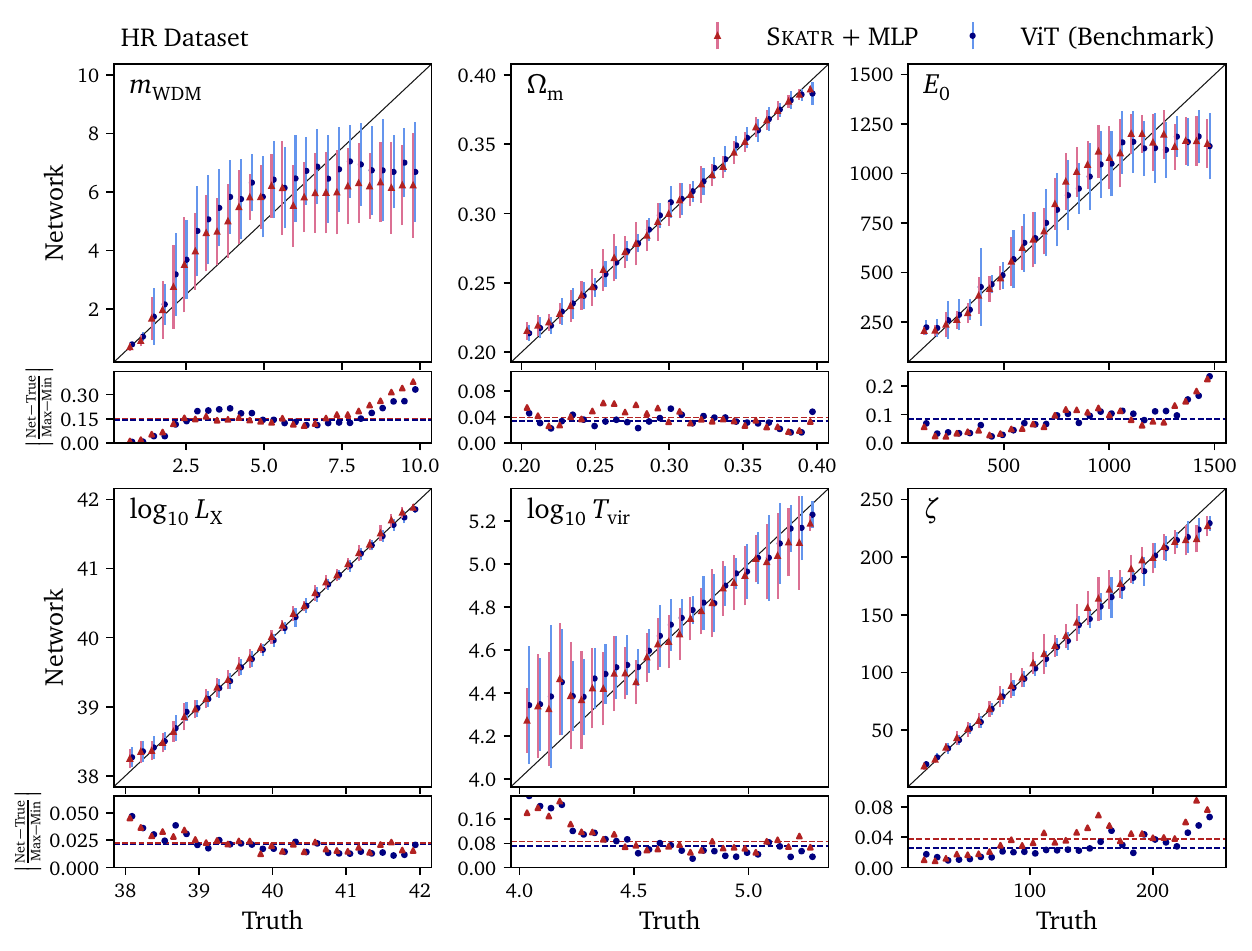}
    \caption{Performance for \textbf{\skatr-summarized LCs, trained
        on the LR dataset, vs ViT trained on the HR dataset}, \skatr is complemented with a 2-layer MLP trained on the HR
      dataset with LCs upsampled to high resolution at
      each iteration.}
    \label{fig:scratch_vs_skatr_hr_upsamp}
\end{figure}

Taking the opposite approach, one can use patches with equivalent
physical size to those used in pre-training, but containing more
voxels. Now, the computational bottleneck shifts to a new embedding
layer, which maps the larger patches into the hidden
dimension of the transformer. While a linear layer would introduce too many parameters,
3D-convolutional layers provide a more efficient
solution.
However, this option uses trainable layers at the input of the network, the transformer backbone must be called at every training iteration, and so there is little efficiency gain.

Alternatively, if the target resolution is known we can pre-train a
\skatr network by upsampling LR LCs at each training iteration. The
advantage of this approach is that one can select a patch size for the
ViTs that suits the target resolution, mitigating computing
bottlenecks. The downside is that pre-training must be run with the
desired resolution in mind. However, the trained \skatr network can be frozen and
used to summarize LCs for a lightweight
MLP. Figure~\ref{fig:scratch_vs_skatr_hr_upsamp} shows the result on
the HR dataset using this setup. In the majority of parameters, the \skatr-MLP combination has average errors on par with the ViT benchmark, with no clear failure modes. However, the performance for $\zeta$ is not matched by the MLP. Here, a loss of precision is evident, with the spread of predictions around the true value being larger for the MLP.

Finally, a combination of the mentioned solutions might work best. The
options that do not repeat \skatr pre-training were not suitable in
our example case primarily due to the large gap in resolution between
LR and HR. A joint approach would be to perform upsampled training at
a set of predefined resolutions then use the other methods to bridge
any remaining resolution difference.

\section{Outlook}
\label{sec:outlook}

The complex structure of 21cm images, combined with the impressive data rate expected at the upcoming Square Kilometer Array presents a new challenge to scientific analyses. Machine learning is our only hope to optimally and completely analyze the SKA dataset. However, the size of training datasets is limited by computational expense of high-resolution simulations, as well as memory requirements.

We present \skatr, a vision transformer (ViT) that learns a highly informative summary of 21cm lightcone data using self-supervision. Our analysis showed that \skatr is capable of leveraging large volumes of relatively cheap data to gain performance on high-resolution simulations.

\textbf{\skatr finds optimal SKA data representations.} Focusing on regression of astrophysical and cosmological parameters, we first established that ViTs are at least as powerful as the previous CNN benchmark (Figure~\ref{fig:cnn-vit-reg-hires}), then demonstrated that \skatr-summarized lightcones contain all information needed to reproduce this performance. In particular, a lightweight MLP trained on frozen \skatr summaries matches the regression accuracy of a full ViT trained from scratch (Figure~\ref{fig:scratch-pretrained-reg}). As a benefit of the \skatr compression, downstream training is extremely cheap, with networks converging over 50 times faster than training from scratch (Figure~\ref{fig:finetune_timing}). Also for simulation-based inference, a generative network conditioned on a fixed \skatr summary yields as constraining posteriors as a jointly-trained ViT (Figures~\ref{fig:likelihoods_calibration_skatr}, \ref{fig:posterior_1d_scratch_vs_pretrain}).

\textbf{\skatr generalizes out-of-domain.} Next we showed that the \skatr summary is also maximally informative out of domain, using datasets simulated at high resolution (HR) and low resolution (LR), with and without noise. The combination of frozen \skatr with a small trainable MLP matches the regression performance of a ViT even in the face of novel parameter correlations and observational noise (Figure~\ref{fig:domain-shift}). This is especially important in radio astronomical observations such as with the SKA, where residual systematics remain and always lead to a data-simulation-gap. Summaries obtained through fully-supervised pre-training did not generalise as well as \skatr and performed worse than a ViTs trained from scratch.

\textbf{\skatr is data efficient.} When considering regimes with limited training data, we find that \skatr scales more favorably than fully-supervised networks (Figure~\ref{fig:data-eff-hr-aug}). \skatr therefore represents a promising solution to data constraints related to high-resolution lightcones, namely that they are expensive to simulate and have a large memory footprint.

\textbf{Resolution adaptation for \skatr.} Finally, we discussed ways to adapt \skatr to the full resolution dataset. While solutions that customize image patching were not viable in our case due to computational expense, we showed that upsampling the LR data during pre-training produces an informative summary.

A number of interesting directions for future work remain. First, the advantages that \skatr offers in terms of compression can be studied further. At high resolution, even a small dataset of 5k lightcones amounts to almost 1TB of disk space. Large datasets of $O(100k)$ lightcones are therefore unlikely to fit in memory. When training a network from scratch, this will limit the amount of data that can be used. However, \skatr could be trained on some fraction of the data and used to compress all available lightcones, allowing a small network to leverage the entire summarized dataset. It would then be interesting to determine whether the large dataset retains its statistical power after being summarized by a network that was trained on a small amount of data.

Secondly, our noise model assumed thermal and instrumental noise as well as a foreground avoidance strategy based on the 21cm foreground wedge. Successful generalization for noise models that include further effects such as radio frequency interference and foreground residuals in the EoR window remains to be shown. Given our model successfully transferred between our noised and noiseless scenarios, we expect \skatr representations to remain informative.

Thirdly, the generalization of \skatr representations to new parameters can be further studied in future work, e.g. to lightcones simulated outside of the prior range. An ambitious limiting case is to use a single cosmology for the pre-training, sampling different initial conditions. The question is then whether inference based on this summary remains sensitive to newly varied parameters. Success in this scenario would allow pre-training on SKA observations, providing an implicit bias to combat large uncertainties in simulations.

\subsection*{Code availability}

The code used for this paper can be found at
\url{https://github.com/heidelberg-hepml/skatr}.

\subsection*{Acknowledgements}

We would like to thank Aaron Nordmann for contributing to this study
in the first phase.  CH's work is funded by the Volkswagen Foundation. CH also acknowledges funding by the Daimler and Benz Foundation.
This work was supported by the 
by the DFG under grant
396021762 -- TRR~257: \textsl{Particle Physics Phenomenology after the
  Higgs Discovery}, and through Germany's Excellence Strategy
EXC~2181/1 -- 390900948 (the \textsl{Heidelberg STRUCTURES Excellence
  Cluster}). It was additionally supported by the Federal Ministry for Education and Science (BMBF) and the Ministry of Science Baden-Wuerttemberg through Germany's Excellence Strategy.

\clearpage
\appendix

\section{Lightcone degeneracy in \texorpdfstring{\mWDM}{mWDM} and \texorpdfstring{\Tvir}{Tvir}}
\label{app:degeneracy}

\begin{figure}[b!]
    \centering
    \includegraphics[width=0.68\textwidth]{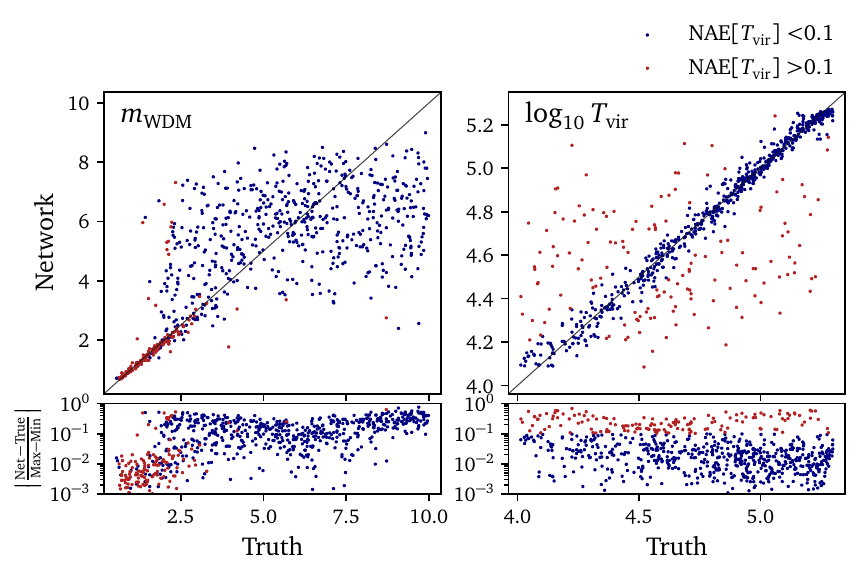}
    \caption{Regression results for a ViT trained from scratch on the HRDS dataset. Points are colored according to absolute error in \Tvir.}
    \label{fig:degeneracy_scatter}
\end{figure}
\begin{figure}[b!]
    \begin{center} 
        \includegraphics[page=2, width=0.39\textwidth]{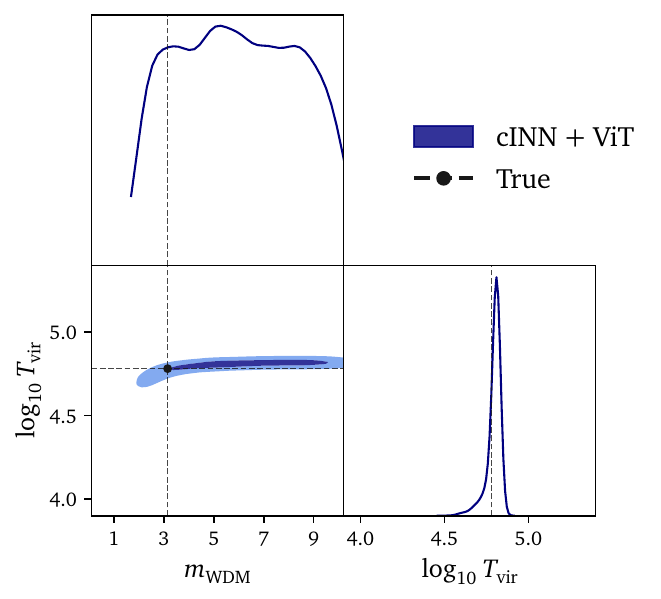}
        \includegraphics[page=4, width=0.39\textwidth]{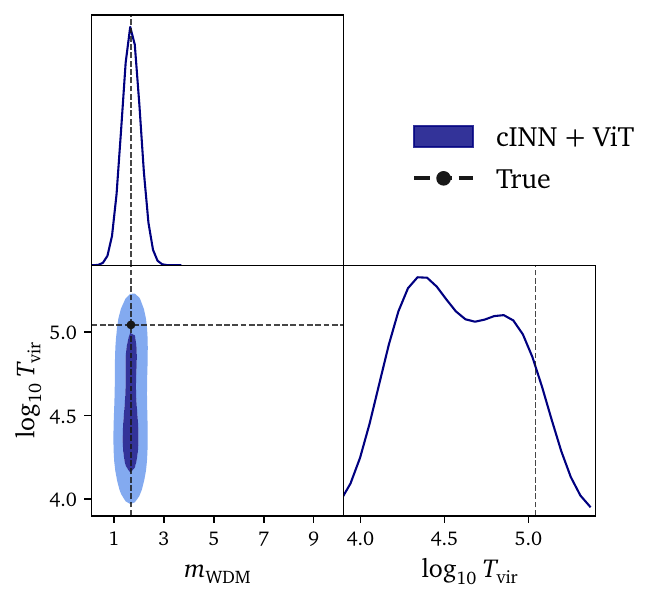}
    \end{center}
    \caption{Parameter degeneracy between \mWDM and \Tvir reflected in the posterior distributions learned by a cINN + ViT in the HRDS dataset. The test LC in the left plot has $\mWDM>2.5\,$keV and the right plot has $\mWDM<2.5\,$keV.}
    \label{fig:posterior_mwdm_Tvir}
\end{figure}

As discussed in the main text, HR-simulated LCs exhibit a degeneracy in the warm dark matter mass, \mWDM, and the minimal virial
temperature, \Tvir. Both parameters bound early star formation and when the limit from \mWDM is stronger than that from \Tvir, then no information on the latter
is available. Here we demonstrate this explicitly using
regression and inference results. Figure~\ref{fig:degeneracy_scatter}
shows a scatter plot of the predicted and true parameter values by a
ViT regressor on the HRDS dataset. Selecting points based on their
normalized absolute error (NAE) in \Tvir reveals a corresponding
cluster in \mWDM. In particular, the outliers in \Tvir correspond
almost directly to the points with low error in \mWDM. This suggests
that the two parameters are not simultaneously predictable in HRDS
lightcones. Figure~\ref{fig:posterior_mwdm_Tvir} shows that same result
is apparent in the posteriors $q(\mWDM, \Tvir|x)$ learned by a cINN +
ViT combination. On the left of the figure is a posterior for a test
point with $\mWDM>2.5\,$keV. Here, the posterior tightly constrains \Tvir,
but has almost maximal uncertainty in \mWDM. On the right side of the
figure, the converse behavior is
observed for an LC with $\mWDM<2.5\,$keV.

\section{Further training details and hyperparameters}
\label{app:training-details}

\begin{table}[tp]
    \centering
    \begin{small} \begin{tabular}{lcc} 
                          & Encoders & Predictor \\
                          \toprule
        Patch size        &  (4, 4, 10) & -  \\
        Embedding dim     & 360  & 48 \\
        Attention heads   &  6   &  4 \\
        MLP hidden dim    & 720  & 96 \\
        Blocks            &  6   &  4 \\
        Positional encoding &  Learnable sin/cos & Fixed sin/cos \\
        Parameters & 6.3M & 110k \\
        \midrule
        EMA rate $\tau$   & \multicolumn{2}{c}{0.9997} \\
        \midrule
        Learning rate
        schedule          & \multicolumn{2}{c}{OneCycle} \\
        Max learning rate & \multicolumn{2}{c}{0.001} \\
        Epochs            & \multicolumn{2}{c}{1000} \\
        Batch size        & \multicolumn{2}{c}{64} \\
        Optimizer         & \multicolumn{2}{c}{AdamW} \\
        Weight decay      & \multicolumn{2}{c}{0.001} \\        
        \bottomrule
    \end{tabular} \end{small}
    \caption{Network and optimization hyperparameters used in \skatr pre-training, discussed in  Section~\ref{sec:skatr}.}
    \label{tab:hyperparams-skatr}
\end{table}

\subsection*{Pre-training}
\label{app:pretrianing-details}

A key component of the \skatr pre-training loss in Eq.~\eqref{eq:pretrain_loss} is the mask sampling procedure, appearing as $p_\text{mask}(M)$. For this sampling, we follow the strategy outlined for video in Ref.~\cite{bardes2024revisiting}. This involves a combination of "long-range" and "short-range" masks, which both span the full redshift (time) dimension of the LC, but have different spatial structure. Long range masks are constructed by sampling three rectangles, each with an aspect ratio in the range [0.75, 1.5] and 70\% coverage of the spatial area, then taking their union. Short range masks are sampled similarly, but using eight rectangles with 15\% coverage. Again, the masks extend across the entire redshift axis.

In the predictor network $h_\psi$ it is not necessary to apply a patching step. This is because its input is already a set of embeddings---those of the context patches. In order to make predictions at the target locations, a set of "mask tokens" are added to the input set. These mask tokens are constructed by summing a shared learnable vector with the positional encoding for each target patch. The full set of patches (context and mask) are then embedded into the hidden dimension of the predictor using a shared linear layer. Similarly, at the predictor output another linear layer projects into the embedding dimension of the target patches.

In order to improve efficiency in training, multiple masks can be sampled per LC. The loss is then calculated for each mask and averaged before taking a gradient step. This saves one evaluation of the context encoder per additional mask. In \skatr, we sample two long-range and two short-range masks per LC.

In Table~\ref{tab:hyperparams-skatr}, we give the full list of hyperparameters for the \skatr networks presented in this paper. With a single NVIDIA H100 GPU, pre-training takes about 50 hours and uses 45GB of GPU memory at batch size 64.

\subsection*{Regression}
\label{app:regression-details}

Table~\ref{tab:hyperparams-reg} lists the hyperparameters we use when training ViTs for regression, including when pre-training for the result in Figure~\ref{fig:data-eff-hr-aug}. For MLPs trained on top of pre-trained summary networks, a faster learning rate of $5\cdot10^{-4}$ is used. Aside from this, the optimization hyperparameters are shared with the ViT. Similarly, the CNN in Figure~\ref{fig:cnn-vit-reg-hires} is trained using the same optimization settings, but a learning rate of $3\cdot10^{-4}$. Its architecture matches exactly the description in Ref.~\cite{Neutsch:2022hmv}. Training times for regression were measured using a single NVIDIA A30 GPU.

Note that the ViT architecture used in \skatr pre-training is larger than the ViTs trained from scratch, with two extra blocks and a wider embedding dimension. We found that training from scratch with the larger network resulted in overfitting and thereby reduced performance. Using a smaller ViT for \skatr degraded performance slightly, but does not strongly affect the results. We understand the lack of overfitting when pre-training to be a consequence of the fact that the loss is based on masking. This means that a single lightcone can provide multiple distinct training objectives, effectively increasing the data efficiency.

\begin{table}[t!]
    \centering
    \begin{small} \begin{tabular}{lcc} 
    \toprule

                          & LR/HRDS           & HR \\
        \midrule 
        Patch size        & (4, 4, 10)    & (7, 7, 50) \\
        Embedding dim     & 144           & 96 \\
        MLP hidden dim    & 288           & 192 \\
        Attention heads   &  \multicolumn{2}{c}{4} \\
        Blocks            &  \multicolumn{2}{c}{4} \\
        Positional encoding &  \multicolumn{2}{c}{Learnable sin/cos }\\
        Patch aggregation &  \multicolumn{2}{c}{Mean} \\
        Head network      &  \multicolumn{2}{c}{MLP} \\
        Parameters & 690k & 540k\\
        \midrule
        Loss              & \multicolumn{2}{c}{Mean Absolute Error} \\
        Learning rate
        schedule          & \multicolumn{2}{c}{Constant} \\
        Learning rate     & $10^{-4}$ & $3\cdot10^{-4}$ \\
        Epochs            & \multicolumn{2}{c}{1000} \\
        Patience          & \multicolumn{2}{c}{50} \\
        Batch size        & \multicolumn{2}{c}{32} \\
        Optimizer         & \multicolumn{2}{c}{AdamW} \\
        Weight decay      & \multicolumn{2}{c}{$10^{-3}$} \\
        \bottomrule
    \end{tabular} \end{small}
    \caption{Network and optimization hyperparameters used to train ViTs for regression. The MLP head architecture is given in Eq.~\eqref{eq:mlp_head}.}
    \label{tab:hyperparams-reg}
\end{table}

\subsection*{Inference}
\label{app:inference-details}
Table~\ref{tab:hyperparams-inf} lists the hyperparameters we use when training cINNs for posterior estimation. ViTs trained as summary networks share the same architecture as for regression (Table~\ref{tab:hyperparams-reg}), except that no head network is used.

\begin{table}[t!]
    \centering
    \begin{small} \begin{tabular}{lc} 
    \toprule
        Bijector            & Rational quadratic spline \\
        Spline bound        & [-10, 10] \\
        Spline bins         & 10 \\
        Block type          & Coupling \\
        Blocks              & 6 \\
        Layers per block    & 2 \\
        Layer dim           & 128 \\
        Channel mixing      & Fixed rotation\\
        Latent distribution & Unit Gaussian\\
        Parameters & 350k\\
        \midrule
        Learning rate
        schedule          & Constant\\
        Learning rate     & $10^{-4}$ \\
        Epochs            & 1200 \\
        Patience          & 100 \\
        Batch size        & 64 \\
        Optimizer         & AdamW \\
        Weight decay      & $10^{-3}$ \\       
        \bottomrule
    \end{tabular} \end{small}
    \caption{Network and optimization hyperparameters used to train cINNs for posterior estimation.}
    \label{tab:hyperparams-inf}
\end{table}

\section{Additional plots}
\begin{figure}
    \centering
    \includegraphics[width=0.65\textwidth]{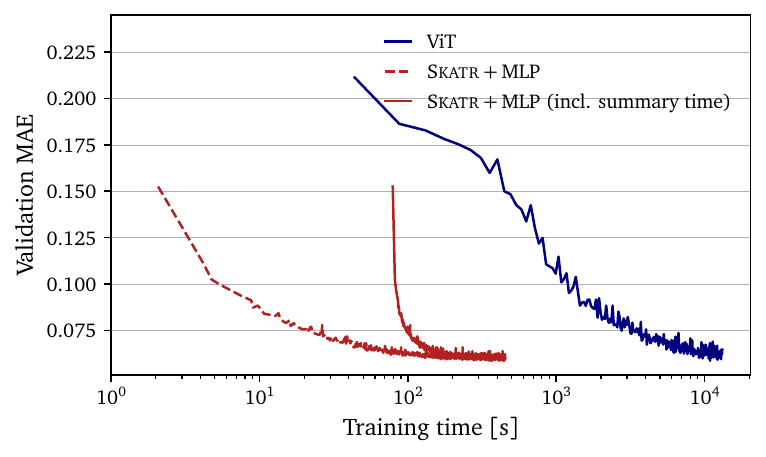}
    \caption{Same as Figure~\ref{fig:finetune_timing}, but for downstream training on the HRDS dataset.}
    \label{fig:finetune_timing_hr}
\end{figure}

\begin{figure}
    \centering
    \includegraphics[width=0.98\textwidth]{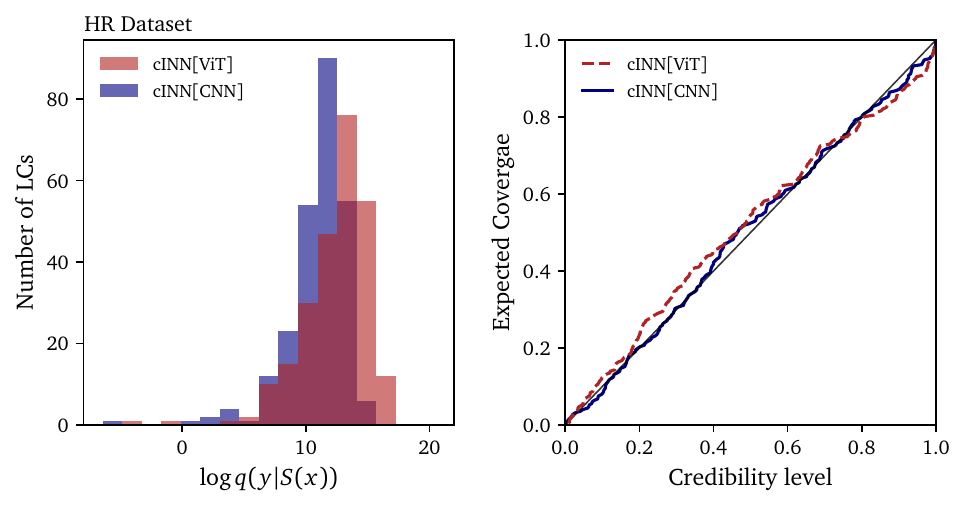}
    \caption{Same as Figure~\ref{fig:likelihoods_calibration_skatr}, but comparing CNN and ViT summaries on the HR dataset.}
    \label{fig:likelihoods_calibration_cnn_vs_vit}
\end{figure}

\begin{figure}
    \centering
    \includegraphics[width=0.98\textwidth]{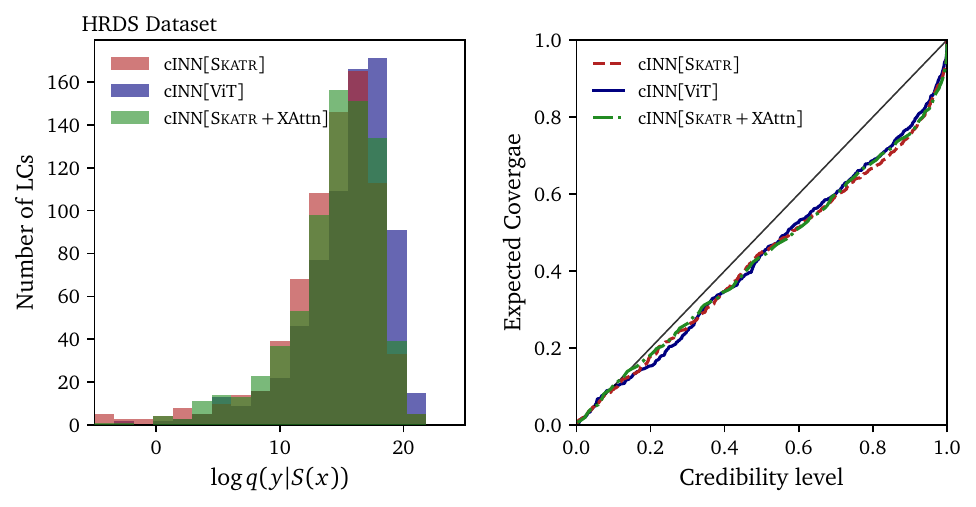}
    \caption{Same as Figure~\ref{fig:likelihoods_calibration_skatr}, but for the HRDS dataset and including a trainable XAttn layer over frozen \skatr embeddings.}
    \label{fig:likelihoods_calibration_skatr_hr_xattn_only}
\end{figure}

\begin{figure}
    \centering
    \includegraphics[width=0.92\textwidth]{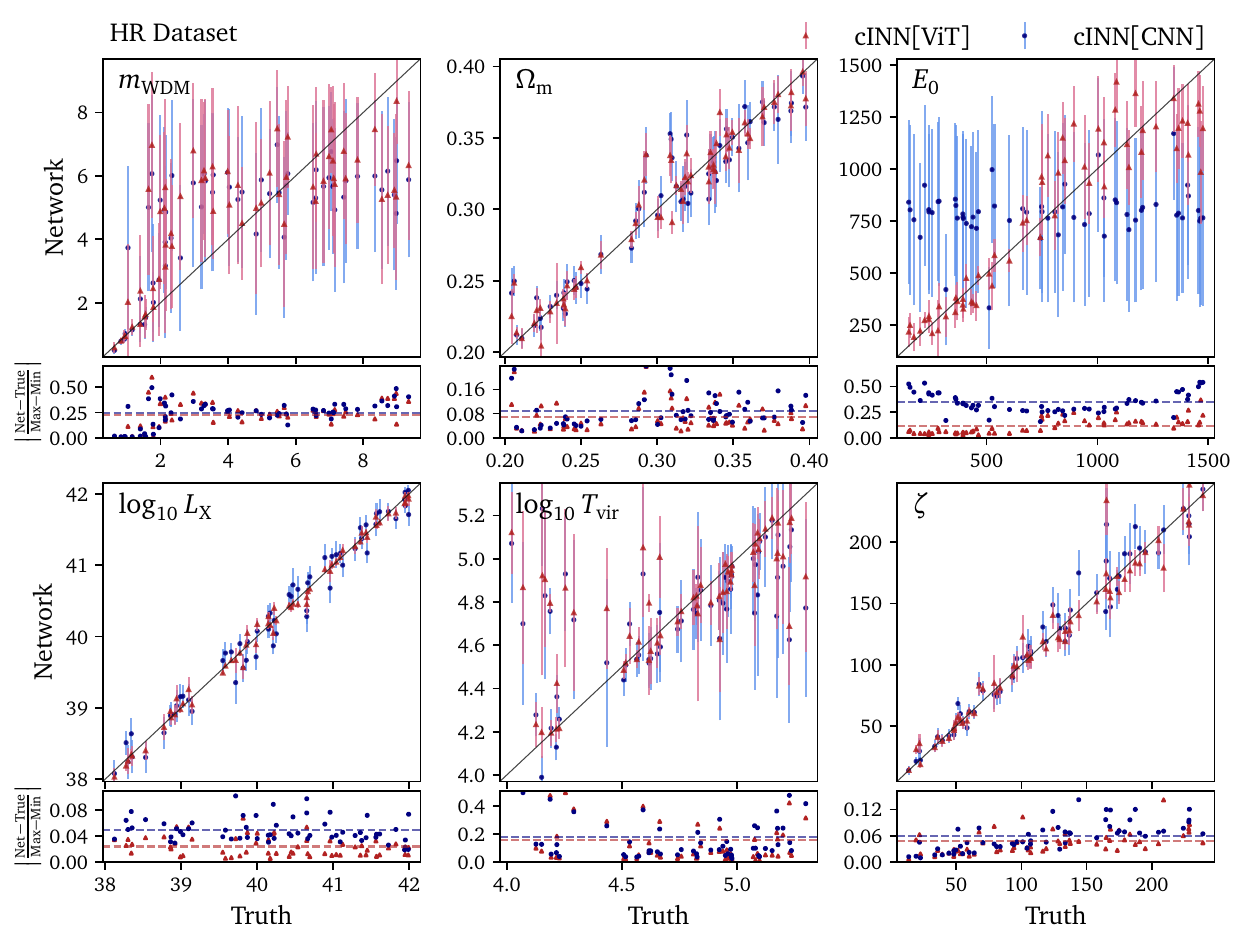}
    \caption{
    Same as Figure~\ref{fig:posterior_1d_scratch_vs_pretrain}, but comparing CNN and ViT summaries on the HR dataset.
    }
    \label{fig:posterior_1d_cnn_vs_vit}
\end{figure}

\begin{figure}
    \centering
    \includegraphics[width=0.92\textwidth]{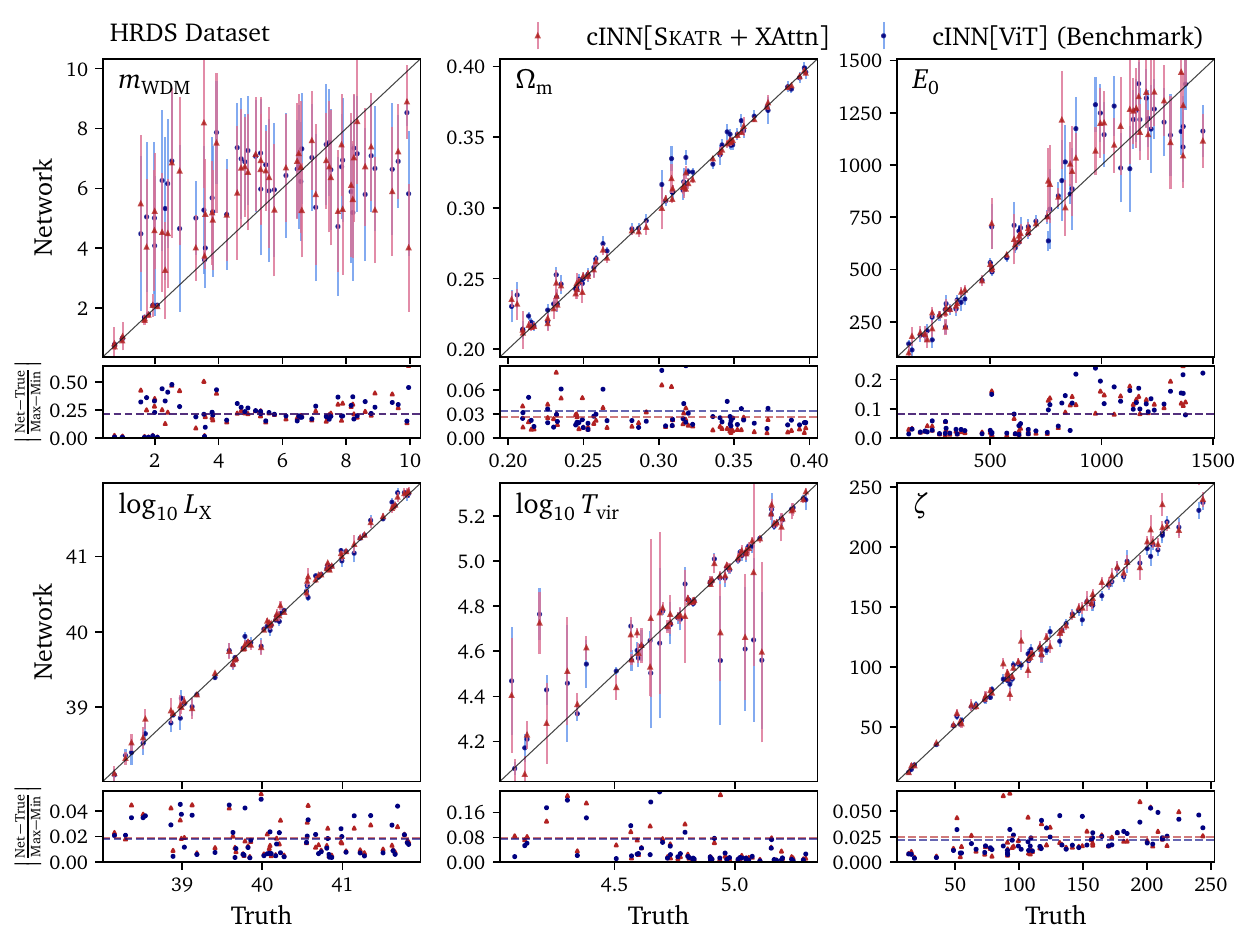}
    \caption{
    Same as Figure~\ref{fig:posterior_1d_scratch_vs_pretrain} but for the HRDS dataset and including a trainable XAttn layer over frozen \skatr embeddings.
    }
    \label{fig:scratch-pretrained-inf-xattn}
\end{figure}

\begin{figure}
    \centering
    \includegraphics[width=\textwidth, page=1]{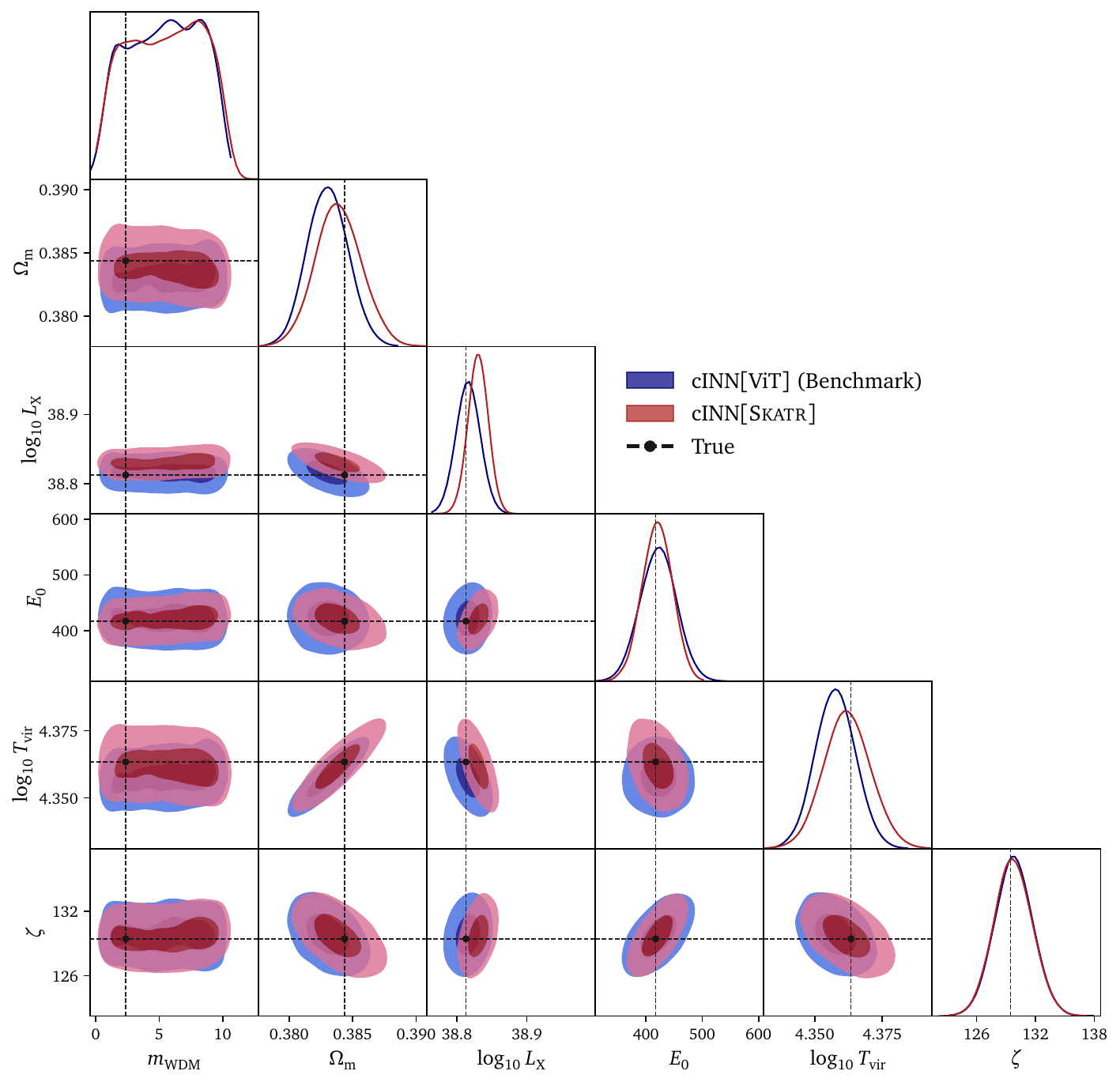}
    \caption{
    Comparison of posterior distributions sampled from a cINN trained jointly on HRDS LCs summarized by either: A complete ViT (blue) or frozen \skatr network (red). The shading levels indicate the 68\% and 95\% highest-density credibility regions.
    }
    \label{fig:scratch-pretrained-inf-lr-1}
\end{figure}

\begin{figure}
    \centering
    \includegraphics[width=\textwidth, page=1]{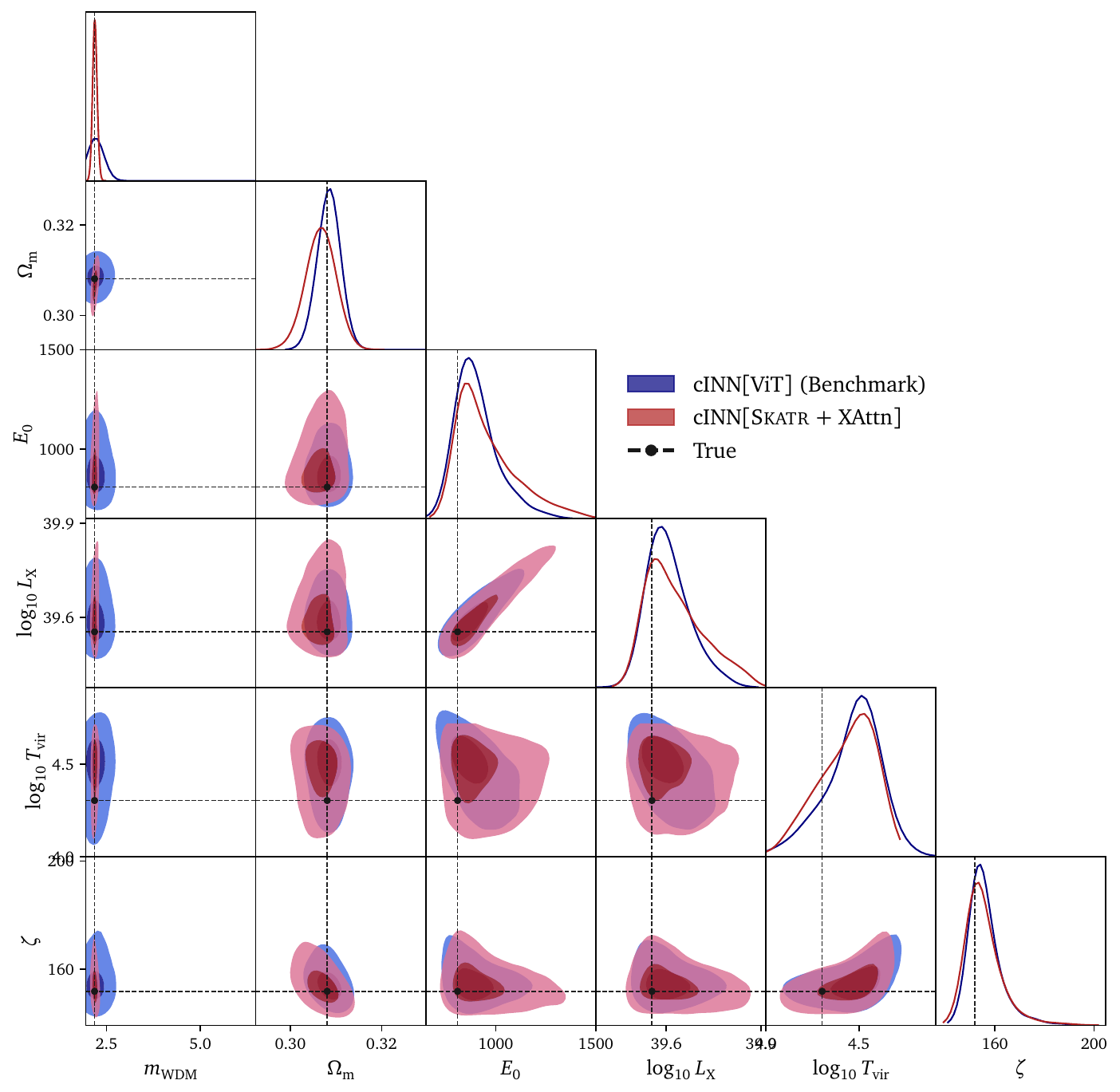}
    \caption{
    Comparison of posterior distributions sampled from a cINN trained jointly on HRDS LCs with either: A complete ViT (blue) or a XAttn pooling over frozen \skatr embeddings (red). The shading levels indicate the 68\% and 95\% highest-density credibility regions.
    }
    \label{fig:scratch-pretrained-inf-1}
\end{figure}

Here we present a selection of supplementary plots:
\begin{itemize}
    \item Training times for downstream regression on the HRDS dataset [Figure~\ref{fig:finetune_timing_hr}]. Similarly to the timing result in the main text, we see more than $50\times$ speed-up in convergence for the MLP compared to training a ViT from scratch. The summarization time is also less significant in this dataset.
    
    \item Posterior likelihoods and calibration:
        \begin{itemize}
            \item CNN vs ViT on HR dataset [Figure~\ref{fig:likelihoods_calibration_cnn_vs_vit}]. The ViT summary network yields higher posterior likelihoods than the CNN on average. Both networks are well calibrated.
            \item ViT vs \skatr with and without XAttn pooling on HRDS dataset [Figure~\ref{fig:likelihoods_calibration_skatr_hr_xattn_only}]. Using the XAttn pooling in \skatr slightly improves the average posterior likelihood, though both approaches are very slightly outperformed by the ViT trained from scratch. The calibration curves are all equally overconfident.
        \end{itemize}

    \item 1D posteriors
        \begin{itemize}
        \item CNN vs ViT on HR dataset [Figure~\ref{fig:posterior_1d_cnn_vs_vit}]. The ViT summary network improves over the CNN in all parameters, with the largest difference in $E_0$.
        \item ViT vs $\operatorname{XAttn}[\skatr]$ on HRDS dataset [Figure~\ref{fig:scratch-pretrained-inf-xattn}]. Using an XAttn pooling to aggregate \skatr summary patches gives marginal posteriors that are equally constraining as the ViT trained from scratch.
        \end{itemize}

    \item 2D posteriors
    \begin{itemize}
        \item ViT vs \skatr on LR dataset [Figure~\ref{fig:scratch-pretrained-inf-lr-1}]. The posteriors from both methods largely agree.
        \item ViT vs $\operatorname{XAttn}[\skatr]$ on HRDS dataset [Figure~\ref{fig:scratch-pretrained-inf-1}]. The posterior from $\operatorname{XAttn}[\skatr]$ constrains $\mWDM$ much more tightly than the ViT, but is slightly wider in the other parameters.
    \end{itemize}
    
\end{itemize}


\clearpage

\clearpage
\bibliographystyle{tepml}
\bibliography{tilman,references}

\end{document}